\documentclass[aip,reprint,fleqn]{revtex4-1}

\usepackage{chemformula} 
\usepackage[T1]{fontenc} 

\usepackage{amsmath,amssymb,amstext}
\usepackage{bbold}
\usepackage{tikz}
\usepackage{array}
\usepackage{booktabs}
\usepackage[parfill]{parskip}
\usepackage{xargs} 
\usepackage{tabularx} 
\usepackage{colortbl} 
\usepackage{float} 
\usepackage{listings}
\usepackage{placeins} 
\usepackage{soul}
\usepackage{siunitx}
\usepackage{hyperref}

\usepackage{xr}
\usepackage{array}
\newcolumntype{C}[1]{>{\centering\arraybackslash}m{#1}}
\newcolumntype{Y}{>{\centering\arraybackslash}X}
\newcolumntype{P}[1]{>{\centering\arraybackslash}p{#1}}

\definecolor{C0}{rgb}{0.12156862745098,0.466666666666667,0.705882352941177}
\definecolor{C1}{rgb}{1,0.498039215686275,0.0549019607843137}
\definecolor{C2}{rgb}{0.172549019607843,0.627450980392157,0.172549019607843}
\definecolor{C3}{rgb}{0.83921568627451,0.152941176470588,0.156862745098039}
\definecolor{C4}{rgb}{0.580392156862745,0.403921568627451,0.741176470588235}
\definecolor{C5}{rgb}{0.549019607843137,0.337254901960784,0.294117647058824}
\definecolor{C6}{rgb}{0.890196078431372,0.466666666666667,0.76078431372549}
\colorlet{C7}{gray!99.607843137254903!black}
\definecolor{C8}{rgb}{0.737254901960784,0.741176470588235,0.133333333333333}
\definecolor{C9}{rgb}{0.0901960784313725,0.745098039215686,0.811764705882353}

\definecolor{dkgreen}{rgb}{0,0.3,0}
\definecolor{dkblue}{rgb}{0,0,0.3}
\definecolor{dkred}{rgb}{0.4,0,0}
\definecolor{gray}{rgb}{0.5,0.5,0.5}
\definecolor{lightgray}{rgb}{0.9,0.9,0.9}
\definecolor{mauve}{rgb}{0.58,0,0.82}
\definecolor{myblue}{rgb}{0,0.03,1}
\definecolor{myred}{rgb}{0.63,0,0}
\definecolor{cyan}{rgb}{0,1,1}
\definecolor{yellow}{rgb}{1,1,0}
\definecolor{orange}{rgb}{1,0.5,0}

\colorlet{cmap1}{dkblue}
\colorlet{cmap2}{cyan}
\colorlet{cmap3}{dkgreen}

\newcommand{\angstrom}{\mbox{\normalfont\AA}}

\newcommand{\energy}[1]{ \pgfmathparse{#1*1} \ensuremath{\num[round-mode=places,round-precision=2]{\pgfmathresult}~eV}} 

\newcommand{\distance}[1]{ \pgfmathparse{#1*1} \ensuremath{\num[round-mode=places,round-precision=2]{\pgfmathresult}~\angstrom}} 

\newcommand{\colorize}[1]{#1}

\newcommand{\PESpoint}{point of the PES}
\newcommand{\PESpoints}{points of the PES}
\newcommand{\predict}{predict}
\newcommand{\prediction}{prediction}
\newcommand{\predicting}{predicting}

\usepackage[colorinlistoftodos, prependcaption, textsize=tiny]{todonotes}
\newcommandx{\LH}[2][1=]{\todo[linecolor=C3,backgroundcolor=C3!25,bordercolor=C3,#1]{\color{C3}\bf LH: #2}}
\newcommandx{\AJ}[2][1=]{\todo[linecolor=C2,backgroundcolor=C2!25,bordercolor=C2,#1]{\color{C2}\bf AJ: #2}}
\newcommandx{\OTH}[2][1=]{\todo[linecolor=C0,backgroundcolor=C0!25,bordercolor=C0,#1]{\color{C0}\bf OTH: #2}}

\emergencystretch=\maxdimen
\begin{document}
	
\author{Lukas H\"ormann}
\author{Andreas Jeindl}
\author{Oliver T. Hofmann}
\email{o.hofmann@tugraz.at}
\affiliation{Institute of Solid State Physics, NAWI Graz, Graz University of Technology, Petersgasse 16, 8010 Graz, Austria}

\title{Reproducibility of Potential Energy Surfaces of \colorize{Organic/Metal} Interfaces on the Example of PTCDA on Ag(111)}

\date{\today}

\begin{abstract}
Molecular adsorption at \colorize{organic/metal} interfaces depends on a range of mechanisms: covalent bonds, charge transfer, Pauli repulsion and van der Waals (vdW) interactions shape the potential energy surface (PES), making it key to understanding \colorize{organic/metal} interfaces.
Describing such interfaces with density functional theory requires carefully selecting the exchange correlation (XC) functional and vdW correction scheme.
To explore the reproducibility of the PES with respect to the choice of method, we present a benchmark of common local, semi-local and non-local XC functionals in combination with various vdW corrections.
We benchmark these methods using perylenetetracarboxylic dianhydride (PTCDA) on Ag(111), one of the most frequently studied \colorize{organic/metal} interfaces.
For each method, we determine the PES using a Gaussian process regression algorithm, which requires only about 50 DFT calculations as input.
This allows a detailed analysis of the PESs' features, such as the positions and energies of minima and saddle points.
Comparing the results from different combinations of XC functionals and vdW corrections enables us to identify trends and differences between the approaches.
PESs for different computation methods are in qualitative agreement, but also displaying significant quantitative differences.
In particular, lateral positions of adsorption geometries agree well with experiment, while adsorption heights, energies and barriers show larger discrepancies.
\end{abstract}


\maketitle 

\begin{quotation}
The following article has been submitted to the Journal of Chemical Physics.
\end{quotation}

\section{Introduction}
\label{sc:Introduction}
Organic/metal interfaces play a vital role in many fields such as microelectronics, heterogeneous catalysis and self assembly.
Studying such interfaces arguably entails investigating their potential energy surfaces (PES), which contain a vast amount of information about the chemistry and physics: 
The global minimum of a PES defines the equilibrium structure at absolute zero temperature, while other local minima represent phases at various thermodynamic conditions.
These conditions in turn depend on the relative depths of the minima, i.e. the adsorption energy of the phases.
The paths and transition rates between these minima depend on the barriers between them, while vibration frequencies of the different structures result from the PESs' curvature.

This wealth of information leads to great interest in theoretically studying PESs, which in turn raises the question of how consistent different approaches are.
Predicting PESs of \colorize{organic/metal} interfaces arguably requires first-principles calculations, which are often some flavor of density functional theory (DFT).
\colorize{While DFT is in principle exact, an exact exchange correlation (XC) functional is still unknown.
Commonly used approximate XC functionals such as local density approximation (LDA) or generalized gradient approximation (GGA) do not fully include van der Waals (vdW) interactions which, however, often play a vital role for the interactions within organic/metal interfaces.\cite{tkatchenko2010van}
VdW interactions arise from non-local fluctuations of the electron density and therefore are not properly described in XC functionals that depend on the local density, such as LDA or GGA.}
Today, many vdW correction schemes exist which remedy this shortcoming.

Studies of interfaces have employed a zoo of different XC functionals and vdW corrections\cite{PhysRevB.76.115421, hauschild2005molecular, sample, gautier2015molecular, ehrlich2011system, maurer2015many} raising the question of reproducibility between different calculation methods.
While several studies exist that investigate the reproducibility of the results of different functionals and vdW corrections, these focus on simple systems, mostly solids or molecules.\cite{yang2010tests, wodrich2007accurate, schultz2005databases, harvey2006accuracy, zheng2009dbh24, lejaeghere2016reproducibility}
Here, we explore the performance for \colorize{organic/metal} interfaces, which are practically more relevant and tend to exhibit more diverse physics.
Specifically, we focus on two questions:
Which combinations of computationally feasible XC functionals and vdW corrections reproduce the experiment most accurately? 
How comparable are PESs of different approaches and will first-principle calculations yield qualitatively, or even quantitatively, equivalent results independent of the chosen functional and vdW correction?

The prototypical system of 3,4,6,10-perylene-tetracarboxylic-dianhydride (PTCDA) on Ag(111) perfectly illustrates this issue.
PTCDA strongly interacts with the surface and has been studied using different functionals:
In a study by Ruiz et al., for instance, LDA showed the smallest adsorption height ($\approx2.6~\angstrom$), pristine PBE (i.e., without a vdW correction scheme) resulted in a non-bonded system, and several other functionals yielded adsorption heights varying by as much as $1~\angstrom$.\cite{TSsurf}
Such discrepancies in calculated adsorption heights have been reported for various molecules on coinage metals.\cite{ruiz2016density, liu2015quantitative, reckien2014theoretical, liu2013structure, li2012molecular}

To shine a spotlight on both earlier raised questions, we use different local, semi-local and non-local XC functionals in combination with different vdW correction schemes to investigate the PES of PTCDA on Ag(111), one of the most frequently studied interfaces.
Molecule and substrate show varied interaction mechanisms that involve (partially) covalent bonds as well as interfacial charge transfer, together with a strong impact of vdW interactions.\cite{Romaner_2009}
Experimental studies have found that PTCDA forms a commensurate adlayer on Ag(111) displaying a herringbone arrangement.
The adlayer contains two different adsorption geometries.\cite{PhysRevB.76.115421, kraft2006lateral}
Rohlfing et al.\cite{PhysRevB.76.115421} proposed that both adsorption geometries sit on bridge sites.
The long molecular axis of geometry (A) is aligned with a primitive substrate lattice vector while geometry (B) is rotated by $17^{\circ}$ (see figure \ref{fig:reproducibility_of_adsorption_geometries}).
The fact that PTCDA forms a commensurate structure on Ag(111) implies that molecule-substrate interactions are dominant compared to molecule-molecule interactions.\cite{hooks2001epitaxy}
More precisely, molecule-substrate interactions result in a PES whose minima are separated by large barriers.
These barriers are so large that molecule-molecule interactions cannot expel an adsorbate molecule from its minimum.
Therefore, this study focuses on the molecule-substrate interactions and investigates the PES of a single molecule on the substrate.
This allows investigating the reproducibility of PESs from different methods and looking into their capability to reproduce the experimental adsorption geometries.

We calculate this PES using a number of different types of functionals and vdW corrections, which we briefly introduce in the remainder of the introduction.

The simplest XC functional is the local density approximation\cite{lda} (LDA). 
It depends only on the electron density at a given point in space and is commonly derived from the homogeneous electron gas model.
It calculates the exchange correlation energy of a test electron and a homogeneous electron gas.
As a result, two closed-shell fragments (e.g., two argon atoms) show an energy minimum as function of their distance (i.e., a bonding interaction), even if they should not. 
This effect is usually called ``overbinding'', and has often been and sometimes still is used to (physically incorrect) ``mimic'' vdW interactions at interfaces.\cite{PhysRevB.76.115421,lee2005ab,tournus2005ab}

The first improvement over LDA is the Generalized Gradient Approximation (GGA).
Due to the inclusion of the gradient, these functionals are also often called semi-local.
In this work, we include three GGA incarnations, PW91,\cite{pw91} PBE,\cite{pbe} and revPBE.\cite{revpbe}
For interface simulations, the PW91 functional has seen frequent use, since it recovers parts of the (spurious) binding of LDA.
The PBE functional, which is a non-empirical (but not parameter-free) simplification of PW91, is considered today’s default functional for materials simulations.\cite{burke2012perspective}
PBE generally underbinds, resulting in molecules that would only bind to the surface via vdW interactions not to bind at all.\cite{Romaner_2009, maurer2019advances}
Our results in this work show that at least for PTCDA on Ag(111), this is the case for PW91 and revPBE as well.
The revPBE functional is a reparametrization of PBE which is said to improve the description of chemisorption.\cite{hammer1999improved}

Local and semi-local XC functionals do not capture the physics of vdW interactions, which originate from non-local electron density fluctuations.
A remedy for this exists in form of a multitude of vdW correction schemes.
We concentrate on some of the most widely used vdW corrections, which are Grimme's D3\cite{D3} (here with Becke-Johnson damping\cite{grimme2011effect}), TS\cite{TS} (Tkatchenko-Scheffler method), TS\textsuperscript{surf }\cite{TSsurf} (which is a reparameterization of TS for specifically for surfaces) and the MBD\cite{MBD} dispersion correction.
All of those are post-processing schemes which do not change the electron density determined during the self-consistent field cycle.
D3, TS and TS\textsuperscript{surf} rely on parameters for the atomic polarizabilities, the dispersion coefficients and the vdW radii.
In D3 these parameters reflect the geometry while in TS and TS\textsuperscript{surf} they result from the local electronic surroundings.
These approaches additionally employ a short-range dampening function, for which different types such as Fermi-type or Becke-Johnson dampening exist.
MBD represents the system using a collection of quantum harmonic oscillators centered at the atomic positions.
These oscillators are characterized by polarizabilities arising from the ground state electron density.
These polarizabilities allow constructing the MBD Hamiltonian, which is diagonalized to determine the MBD energy correction.
MDB is said to be more accurate\cite{blood2016analytical, maurer2015many} than D3 and TS but is also significantly more computationally intensive.

Beyond that, non-local XC functionals seek to directly capture non-local interactions.
A commonly used example is vdW-DF,\cite{vdwDF} which is essentially a two-body correlation-correction and hence neglects many body effects.
The correction employs linear response theory using a response function constructed from a plasmon-pole-type approximation to the local polarizability.

The next improvements over GGA-type functionals would be meta-GGA and hybrid functionals.
However, so far meta-GGAs have not seen widespread use for interfaces.
Therefore, here we use only the SCAN-functional, which has recently gained popularity, for our benchmarks.
Hybrid functionals which admix Hartree-Fock-like exchange are frequently discussed for \colorize{organic/metal} interfaces due to their frequent use in chemistry.
\colorize{For semiconductor/organic interfaces, where the level alignment,\cite{xu2013space, wang2019modulation} the amount of localization of charge transfer across interfaces,\cite{hofmann2015integer, gruenewald2015integer} as well as the energy contribution from band bending depends sensitively on the band gap,\cite{moll2013stabilization, sinai2015multiscale} hybrid functionals are indispensable. For metal/organic interfaces, however, usually the electronic properties do not notably improve (or even deteriorate)\cite{fabiano2009towards, biller2011electronic, hofmann2013interface} unless magnetism plays a significant role\cite{wruss2019magnetic}.}
Moreover, choosing the correct amount of exchange is system-dependent and non-straightforward.\cite{wruss2018distinguishing, atalla2013hybrid, hofmann2013interface}
Furthermore, many popular hybrid functionals are intrinsically less suited for metals.\cite{paier2007does}
On more practical grounds, hybrid functionals are typically prohibitively expensive.
Therefore, although some of our earlier results indicate that hybrid functionals may have a profound impact on the adsorption geometry,\cite{wruss2019impact} we do not consider them in this study.

\section{Methodology}
\label{sc:Methodology}

As stated earlier, we study the reproducibility of PESs in relation to the choice of XC functional and vdW correction using the example of PTCDA on Ag(111).
Hereafter, we will refer to combinations of functionals and vdW correction schemes as methods.
This section will first briefly introduce the DFT settings used for the different employed methods.
Since none of these allow brute forcing the PES, we use an algorithm based on Gaussian process regression (GPR), which we discuss in the second half of this section.
Hereby, we focus on the PES of the most important degrees of freedom, rather than on the full PES.

We use the VASP\cite{kresse1993ab, kresse1994ab, kresse1996efficiency, kresse1996efficient} quantum chemistry code with the recommended PAW pseudopotentials\cite{blochl1994projector, kresse1999ultrasoft} and set the energy cutoff to {\energy{800}} for all calculations, ensuring that the total energy of PTCDA is converged to within {\energy{0.05}}.
A comprehensive summary of the convergence tests can be found in section \ref{SIsc:convergence_tests} of the supporting information.
We calculate PESs with the following methods:
First we use \textbf{LDA} because it has been extensively used in the past since it ``mimics'' vdW interactions.
Next come three GGA functionals, namely \textbf{PW91}, \textbf{PBE} and \textbf{revPBE}.
We use the GGAs without vdW correction and in combination with the \textbf{D3}, \textbf{TS} and \textbf{TS\textsuperscript{surf}} correction, allowing us to demonstrate the effect of vdW corrections.
Additionally, we combine PBE with \textbf{MBD} to test a more accurate,\cite{blood2016analytical, maurer2015many} but also more expensive vdW corrections as well.
Finally, we employ the original \textbf{vdW-DF} (implementation by Klimes et al.\cite{klimevs2009chemical, klimevs2011van}) as well as the newer \textbf{rVV10}\cite{peng2016versatile} functional to gauge non-local XC functionals.
The latter is combined with the SCAN meta-GGA functional, allowing us to also study a meta-GGA functional.

We model the interface using the repeated slab approach and a $6 \times 6$ substrate supercell with $6$ Ag layers and $15~\angstrom$ vacuum.
This supercell can readily accommodate one PTCDA molecule in any surface orientation, ensuring that its periodic replica are sufficiently separated such that the intermolecular interactions fall below  {\energy{0.025}} per molecule.
Further, we individually determine the Ag lattice constant for each method (see section \ref{SIsc:substrate_lattice_constant} of the supporting information).
Hereby we use the primitive bulk unit cell with a uniform $\Gamma$-centered $36 \times 36 \times 36$ k-grid and minimize the total energy employing a Birch–Murnaghan fit.
This k-grid results from converging the total energy of the bulk to an uncertainty of $0.0005~eV$ per Ag-atom.
Consequently, we use a $\Gamma$-centered $6 \times 6 \times 1$ k-grid for the interface supercell.

Additional geometric considerations result from the GPR algorithm used to {\predict} the PES (a detailed explanation of the algorithm follows later):
This algorithm determines the PES of a flat rigid molecule on the fixed substrate, using single-point calculations as input data.
The assumption of a flat rigid molecule is justified since PTCDA remains largely planar when adsorbing on Ag(111).
\colorize{
Experimental measurements show\cite{hauschild2005molecular} that the four carboxylic O atoms are displaced approximately $0.18~\angstrom$ out of plane (downward) while the C backbone remains flat.
Conversely, the two anhydride O atoms are approximately $0.11~\angstrom$ above the C backbone.
This distortion from planarity (which is quite small compared to many other adsorbates) results from the formation of a (partially) covalent bond between carbonyl groups and the substrate, which leads to a (partial) rehybridization of the carbon atoms.\cite{PhysRevB.76.115421}
Keeping the molecule flat may raise the caveat of possibly missing or underestimating the covalent bonding part of the PES.
However, as we show below, also within this approximation we readily reproduce the previously reported adsorption geometries (see figure \ref{fig:reproducibility_of_adsorption_geometries}), which shows that this approximation does not have a qualitative impact on the PES.
Also, the agreement with previously reported energies and heights\cite{maurer2015many, TSsurf}, where the molecule was fully relaxed, indicates that keeping it flat is an acceptable approximation.
}

With the DFT-related aspects covered, we can think about the challenge of calculating PESs.
First, the property of interest in our case is the adsorption energy $E_{ads}$.
We define $E_{ads}$, using the total energy of the combined system $E_{mol+sub}$, the energy of a PTCDA molecule in vacuum $E_{mol}$ and the energy of the clean substrate $E_{sub}$.
\begin{equation}
E_{ads} = E_{mol+sub} - E_{mol} - E_{sub}
\end{equation}

\begin{figure}
	\centering
	\begin{minipage}[top]{0.49\linewidth}
		\includegraphics[width=\linewidth]{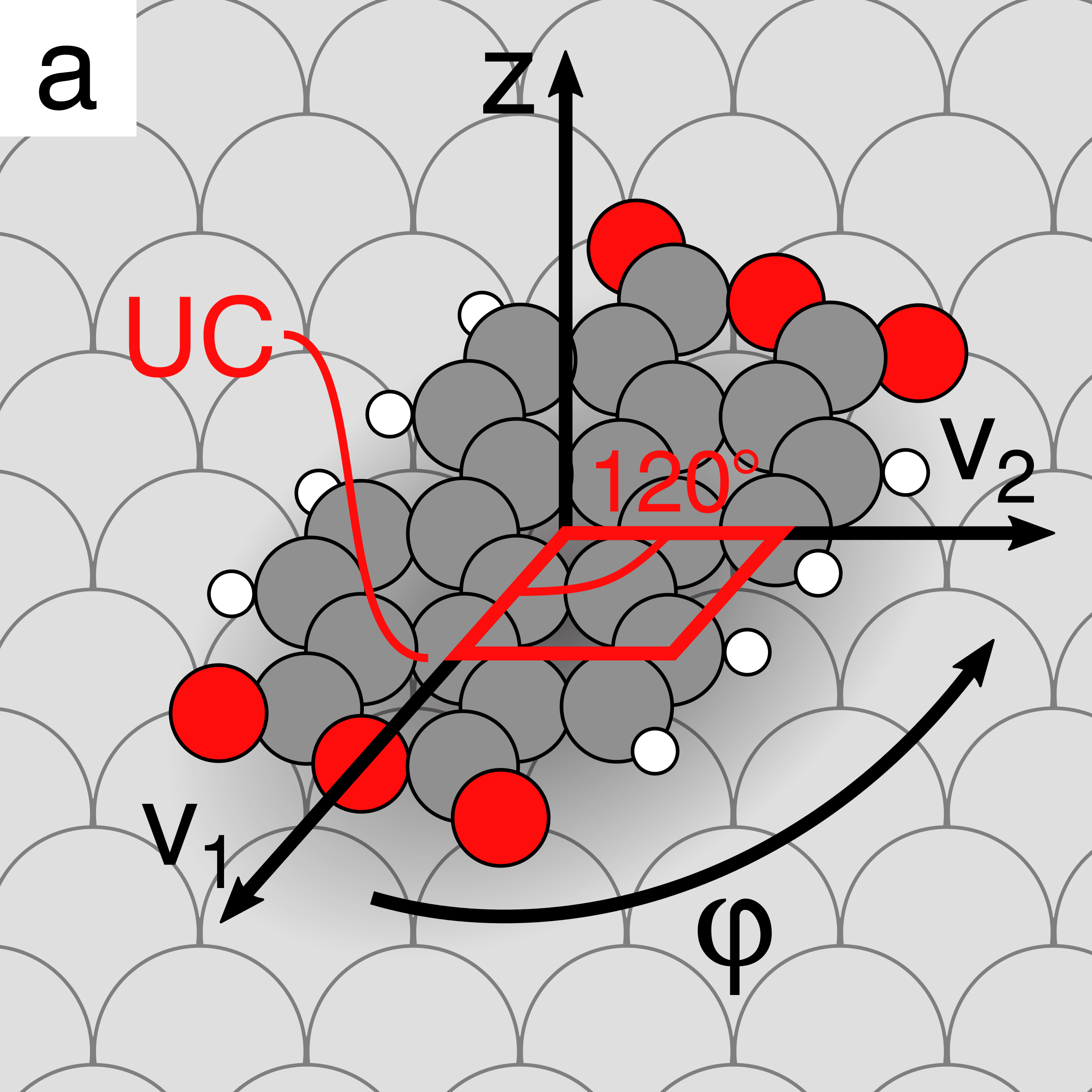}
	\end{minipage}%
	\begin{minipage}[top]{0.02\linewidth}
		\hfill
	\end{minipage}%
	\begin{minipage}[top]{0.49\linewidth}
		\includegraphics[width=\linewidth]{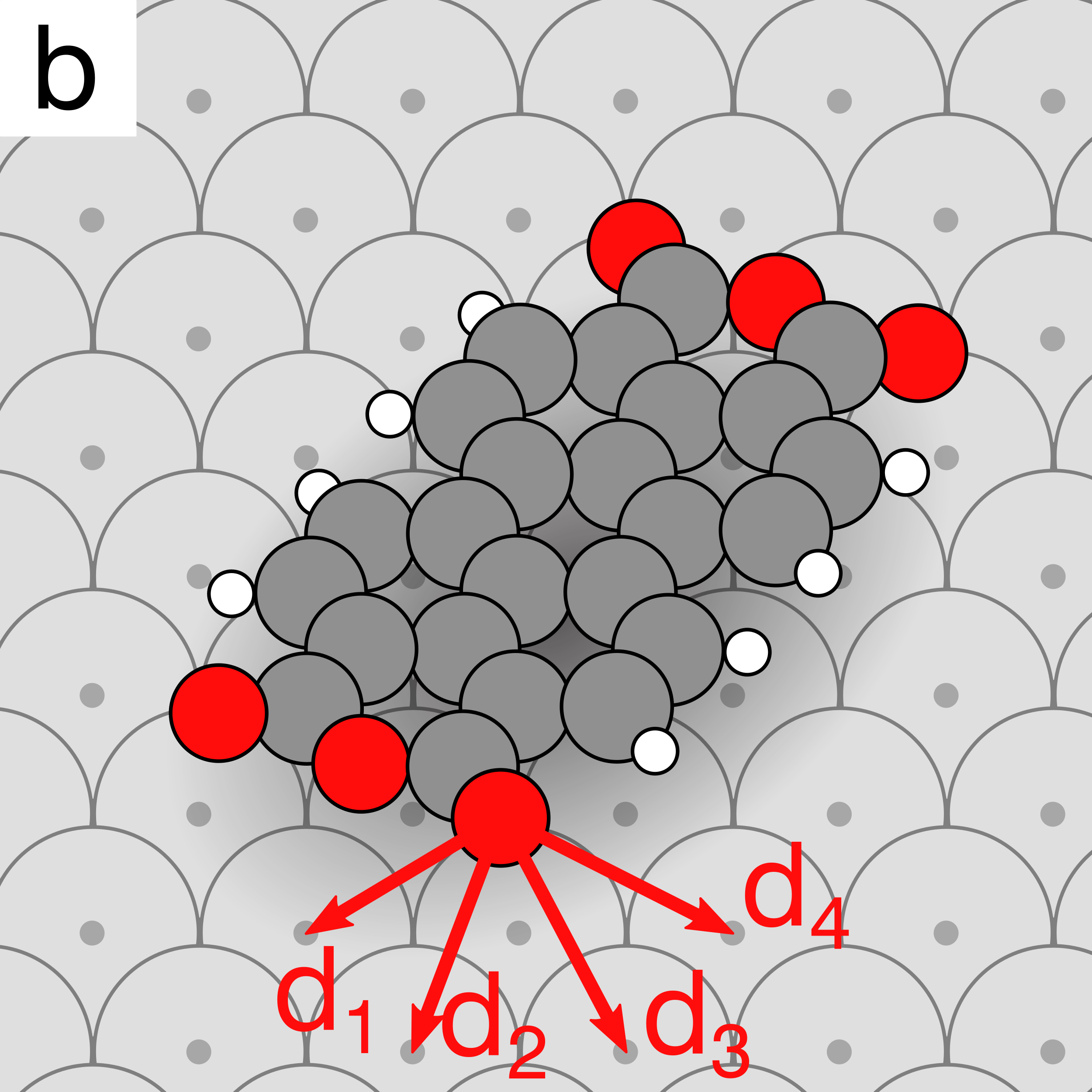}
	\end{minipage}
	\caption{a) Degrees of freedom of the PES, unit cell is indicated in red; b) Illustration of the feature vector $\mathbf{f^\alpha}$ used in this publication; \colorize{The coordinates $v_1=0$, $v_2=0$ correspond to the bridge position and the molecule is rotated by $\phi=0$}}
	\label{fig:feature_vector}
\end{figure}

Determining the PES requires finding the adsorption energy of PTCDA on an Ag(111) surface for a large number of different positions or orientations of the molecule on the substrate. 
In fact, considering all degrees of freedom would render this task computationally intractable.
Therefore, we use a simplified system of a rigid PTCDA molecule on a fixed Ag(111) substrate.
There, we have four degrees of freedom, namely the position of the molecule's center of mass specified via $v_1$, $v_2$, $z$ and the rotation $\phi$ around the $z$ axis (see figure \ref{fig:feature_vector}a).
The $z$ axis is oriented perpendicular to the Ag(111) surface, while $v_1$ and $v_2$ are fractional coordinates of the primitive substrate lattice vectors $\mathbf{v_1}$ and $\mathbf{v_2}$.
These four coordinates span our search space, for which we now need to define boundaries.
Periodic boundary conditions for $v_1$ and $v_2$ are given by the primitive substrate unit cell.
For $\phi$ we use rotation and mirror symmetries of both the molecule and the substrate to limit the search space to $[ -\pi/6, \pi/6 ]$.
The boundary conditions for the adsorption height $z$ are $[2.5, 3.5]~\angstrom$ accounting for the experimental adsorption height of $2.86~\angstrom$\cite{hauschild2005molecular, henze2007vertical} and accommodating previously reported theoretical adsorption heights.\cite{TSsurf}
Having defined boundaries for the search space leads us to choosing points to calculate.
Here, a uniform grid would be a simple and straightforward way.
Using a coarse grid with $10$ divisions in each degree of freedom would already require calculating $10^4$ different adsorption geometries of PTCDA on Ag(111) to determine the PES with a single method --- a task that is clearly intractable with current computing resources.

To overcome this computational bottleneck, we employ a machine-learning algorithm based on GPR.
GPR has proven its usefulness for investigating interfaces with a notable example being the BOSS code.\cite{todorovic2019bayesian, doi:10.1002/advs.202000992}
The present algorithm differs from BOSS in employing a more general descriptor, which we explain below.

For our case, GPR may be understood as a sophisticated method of interpolating adsorption energies.
In other words, we calculate the energies for a number of positions and orientations of the molecule and {\predict} the energies in-between.
Our algorithm uses adsorption energies as well as atomic forces as input data.
To simplify the explanation of GPR we focus only on energies.
Mathematically, we formulate GPR as a conditional probability for the energies we wish to {\predict} $\mathbf{E_{P}}$ given the calculated energies $\mathbf{E_{C}}$.
This probability comes in form of a multivariate Gaussian with a matrix $A$ and the expectation value $\bar{\boldsymbol{\mu}}$:
\begin{equation}
p(\mathbf{E_{P}}|\mathbf{E_{C}}) \propto \exp{ \left[ -\frac{1}{2} ( \mathbf{E_{P}} - \bar{\boldsymbol{\mu}} )^T ~A~ ( \mathbf{E_{P}} - \bar{\boldsymbol{\mu}} ) \right] }
\end{equation}

The best estimator for $\mathbf{E_{P}}$ is the expectation value $\bar{\boldsymbol{\mu}}$ which can be calculated in the following way:
\begin{equation}
\bar{\boldsymbol{\mu}} = \boldsymbol{\mu_P} + C^{PC} ( C^{CC} + \sigma^2 \mathbb{1} )^{-1} (\mathbf{E_C} - \boldsymbol{\mu_C}) 
\end{equation}

The equation for $\bar{\boldsymbol{\mu}}$ requires an uncertainty $\sigma$, a prior mean $\boldsymbol{\mu} = (\boldsymbol{\mu_C}, \boldsymbol{\mu_P})^T$ and a covariance matrix $C$.

$\sigma$ represents the numerical uncertainty of the calculated adsorption energies $\mathbf{E_{C}}$.

The prior mean $\boldsymbol{\mu} = (\boldsymbol{\mu_C}, \boldsymbol{\mu_P})^T$ is the initial estimate for the adsorption energies.
It consists of two parts, $\boldsymbol{\mu_P}$ for the energies to {\predict} and $\boldsymbol{\mu_C}$ for the calculated energies.

In addition to the expectation value $\bar{\boldsymbol{\mu}}$, GPR yields a model uncertainty, which we find in the diagonal elements of the matrix $A$.
We calculate the matrix $A$ from the uncertainty $\sigma$ and the covariance matrix $C$.
\begin{equation}
\label{eq:matrix_A}
A^{-1} = C^{PP} + \sigma^2 \mathbb{1} - C^{PC} ( C^{CC} + \sigma^2 \mathbb{1} )^{-1} C^{CP}
\end{equation}

The most important component of a GPR is arguably the covariance matrix $C$.
Each of its elements $C_{\alpha\beta}$ serves as a measure of similarity for two {\PESpoints}, with a {\PESpoint} referring to an adsorption geometry and its associated energy ($v_1$, $v_2$, $\phi$, $z$, $E_{ads}$).
The covariance matrix $C$ consists of four sub matrices $C^{PP}$, $C^{PC}$, $C^{CP}$ and $C^{CC}$, whereby $(C^{PC})^T = C^{CP}$.
$C^{PP}$ is the covariance matrix between {\PESpoints} we {\predict}.
$C^{PC}$ is the covariance matrix between points we {\predict} and points we calculate. 
$C^{CC}$ is the covariance matrix between points we calculate.
The measure of similarity $C_{\alpha\beta}$ between two {\PESpoints} $\alpha$ and $\beta$ depends on the similarity between the respective geometries.
For this purpose we represent each geometry using a feature vector $\mathbf{f^{\alpha}}$ containing $N^\alpha$ elements.
We construct this feature vector from inverse distances $d_i$ between the respective atoms of adsorbate and substrate.

\begin{equation}
\label{eq:feature_vector}
\mathbf{f^\alpha} = \left( \dots, \left( \frac{d_i}{d_{min}} \right)^n, \dots \right)
\end{equation}

Here $d_{min}$ and $n$ are hyperparameters.
$d_{min}$ is a distance threshold, where we consider atoms with a smaller pair distance as colliding.
$n$ is a decay power set to negative integer values.

We calculate an element of the covariance matrix using two feature vectors $\mathbf{f^{\alpha}}$ and $\mathbf{f^{\beta}}$ in the following way:
\begin{equation}
\label{eq:kernel}
C_{\alpha\beta} \propto \sigma^2_k \sum_{i = 1}^{N^\alpha} \sum_{j = 1}^{N^\beta} ~\exp \left[ -\frac{(f^\alpha_i - f^\beta_j)^2}{4 \tau^2} \right] ~g(f^\alpha_i) ~g(f^\beta_j)
\end{equation}

Here $\tau$ and $\sigma_k$ are hyperparameters.
$\tau$ is the feature decay length and $\sigma_k$ is a global scaling factor for the similarity.
We normalize the covariance matrix such that $\sqrt{ C_{\alpha\alpha} } = \sigma^2_k$.
$g(f_i^{\alpha})$ is a decay function (see supporting information \ref{SIsc:gaussian_process_regression}).
$f_i^{\alpha}$ are elements of the feature vector $\mathbf{f}^{\alpha}$, which is a suitable mathematical representation of a particular adsorption geometry.

For some hyperparameters, such as $d_{min}$, $d_{max}$ and $n$ we use physically motivated settings (see  section \ref{SIssc:hyperparameters_in_gaussian_process_regression} of the supporting information).
However, this cannot be done for the feature decay length $\tau$ and the global scaling factor $\sigma_k$.
Hence, we optimize these two hyperparameters by minimizing the marginal likelihood.

Next, the GPR algorithm needs a supply of training data.
We select training points by determining positions with the highest model uncertainty of the GPR model (see equation \ref{eq:matrix_A}).
For the sake of consistency, we use the same {\PESpoints} as training data for all methods.
At each of these positions we calculate the energy and its derivatives with respect to the position (i.e., the forces) for every method. 
We use a total of $49$ symmetry-inequivalent {\PESpoints}, which allows obtaining GPR uncertainties below  {\energy{0.040}} ($1~kcal/mol$) (see section \ref{SIsc:prediction_uncertaintiy} of the supporting information).

Finally, we use the symmetries of the substrate to generate symmetry equivalent {\PESpoints}, thereby multiplying the number of data points.
These symmetry equivalent data points provide both the adsorption energy and the atomic forces as input data for the GPR algorithm.
We note that forces for revPBE with TS or TS\textsuperscript{surf} do not improve the {\prediction} accuracy and are not used as input data when {\predicting} PESs for these two methods.
Still, in these cases the {\prediction} uncertainty remains below  {\energy{0.040}} (see section \ref{SIsc:prediction_uncertaintiy} of the supporting information).

\section{Results and Discussion}
\label{sc:Results_and_Discussion}

This section will first compare qualitative features of the PESs and then focus on investigating quantitative differences and similarities between the different methods.
Hereby, we look into the reproducibility of theoretical results regarding adsorption heights and geometries and compare these results to experiment.
Throughout this section, we use the PBE functional and the TS\textsuperscript{surf} correction as reference method, since it shows the best agreement with the experimental adsorption geometries.

\begin{figure}[H]
	\centering
	\includegraphics[width=\linewidth]{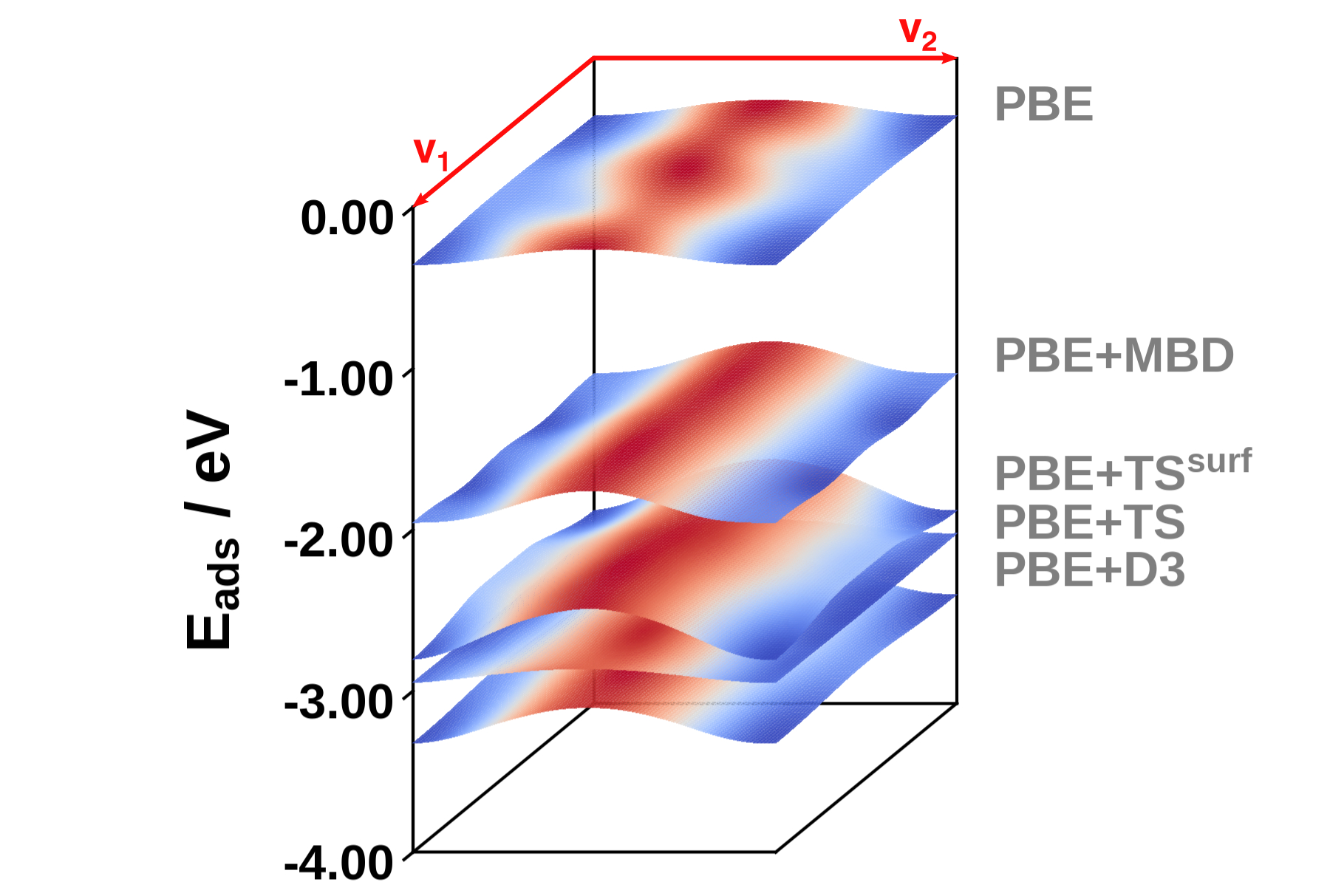}
	\caption{PESs of PBE with different vdW corrections; $\mathbf{v_1}$ and $\mathbf{v_2}$ are the primitive substrate lattice vectors}
	\label{fig:pes_3d_pbe}
\end{figure}

First, we wish to highlight a number of qualitative trends and offer a quantitative discussion later in the results section.
The qualitative agreement between the PESs of the different methods is exemplarily shown for the PBE functional with various vdW corrections in figure \ref{fig:pes_3d_pbe} (section \ref{SIsec:potential_energy_surfaces} of the supporting information shows the PESs of all methods).
For the sake of illustration, we only depict a two-dimensional cross section of the PES, where we show the adsorption energy of PTCDA for different $v_1$ and $v_2$ positions in the primitive substrate unit cell (see figure \ref{fig:feature_vector}a).
$\phi$ is fixed such that the long axis of PTCDA aligns with a primitive substrate lattice vector.
This is the same orientation that PTCDA has in the experimentally proposed adsorption geometry (A).
$z$ is optimized to its minimum energy at every point, i.e. the molecule assumes the optimal adsorption height at every lateral position.
Overall, energetically favorable and unfavorable parts of PESs from different methods lie in the same position.
For instance, we find small barriers when translating the molecule in direction of the lattice vector $\mathbf{v_1}$ and significantly higher barriers for translation in $\mathbf{v_2}$.
Physically, this implies that a PTCDA molecule could move easily in one direction, but not the other --- a conclusion which would be obtained with any of the methods used in this work.

\begin{figure}[h]
	\centering
	\includegraphics[trim={0 1.2cm 0 0}, clip,width=\linewidth]{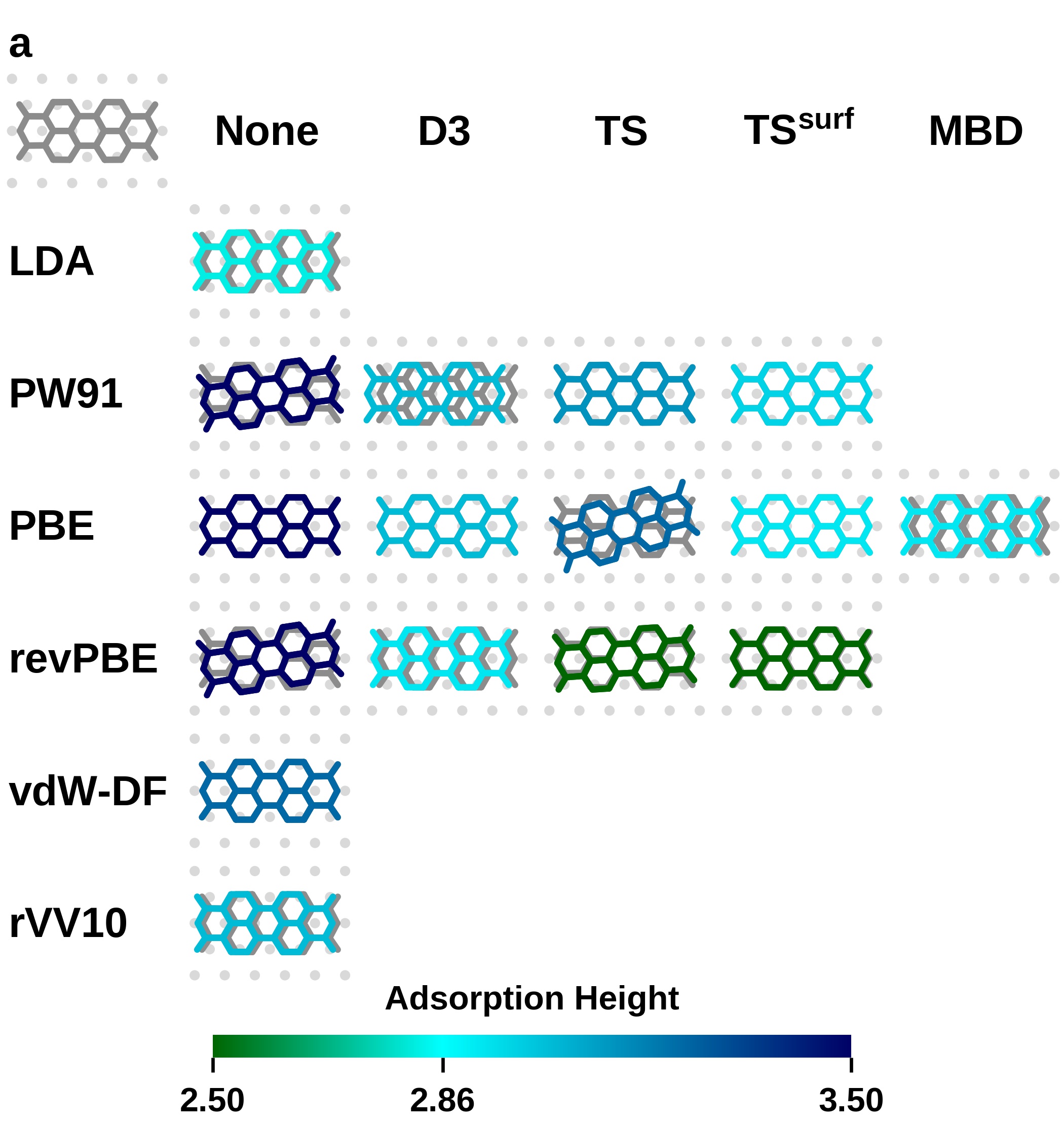}
	\vspace{0.0cm}
	
	\includegraphics[width=\linewidth]{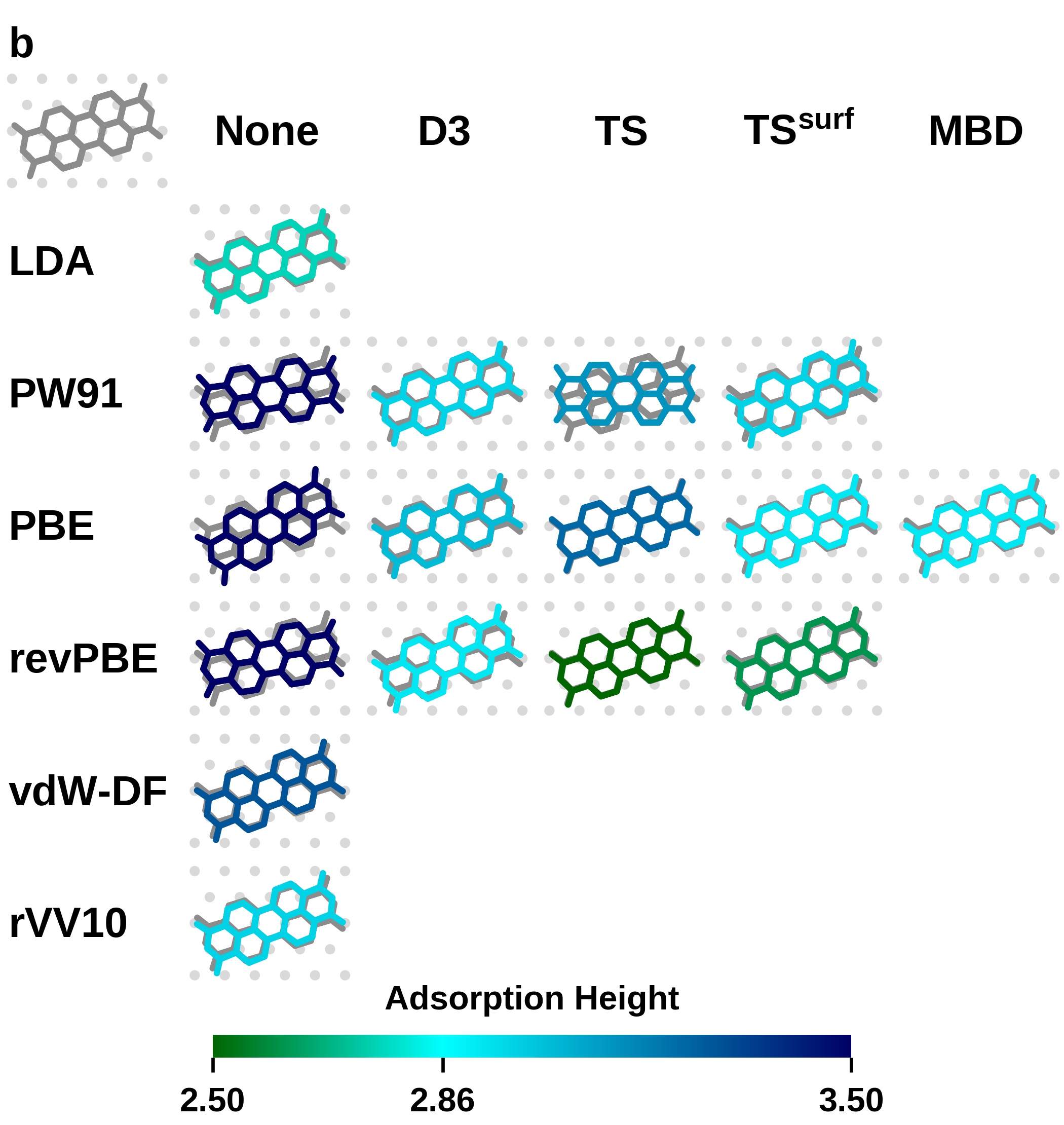}
	\caption{Reproducibility of the experimentally proposed adsorption geometries\cite{PhysRevB.76.115421, kraft2006lateral, hauschild2005molecular, henze2007vertical} displayed in the upper left corner of each table and as the shadow below the colored molecules; color indicates adsorption height; a) experimental adsorption geometry (A); b) experimental adsorption geometry (B)}
	\label{fig:reproducibility_of_adsorption_geometries}
\end{figure}

Despite the qualitative agreement, we observe significant quantitative discrepancies in adsorption energy and transition barriers.
Differences in absorption energy can be highlighted by using the example of experimental adsorption geometry (B).
Most methods that include vdW interactions yield adsorption energies between {\energy{-2.00}} and {\energy{-3.00}} scattering around the expected adsorption energy of $-2.40~eV$.\cite{tkatchenko2010van}
The outliers are \colorize{PBE+MBD with the smallest ({\energy{-1.99}})} and revPBE+TS with the largest absorption energy ({\energy{-5.71}}) (see tables \ref{SItab:adsorption_heights_and_rankings_of_minimum_A} and \ref{SItab:adsorption_heights_and_rankings_of_minimum_B} in the supporting information).
Similar discrepancies exist for barrier heights, where we consider the barrier between the minima most closely resembling the two experimental geometries (see figure \ref{fig:reproducibility_of_adsorption_geometries}).
VdW-DF yields the smallest ($\Delta E^{AB}=${\energy{0.011}}) and \colorize{LDA the largest barrier ($\Delta E^{AB}=${\energy{0.082}}).}
However, not all methods yield the same minima, making a systematic comparison difficult.
Therefore, we investigate the corrugation of the adsorption energy, which allows to instructively illustrate the origin of these discrepancies.
We define the corrugation as the difference between the highest and lowest energy of two-dimensional ($v_1$, $v_2$) cross sections of the PES.
Hereby, we focus on the individual contributions of the functional and the vdW correction to the corrugation before highlighting quantitative differences.
For methods with a posteriori corrections (D3, TS, TS\textsuperscript{surf}, MBD), we can straightforwardly separate the contributions of the functional and the vdW correction.
The vdW correction shows only low corrugation, becoming nearly constant at adsorption heights above {\distance{3.00}}.

\begin{figure}[h]
	\centering
	\includegraphics[width=\linewidth]{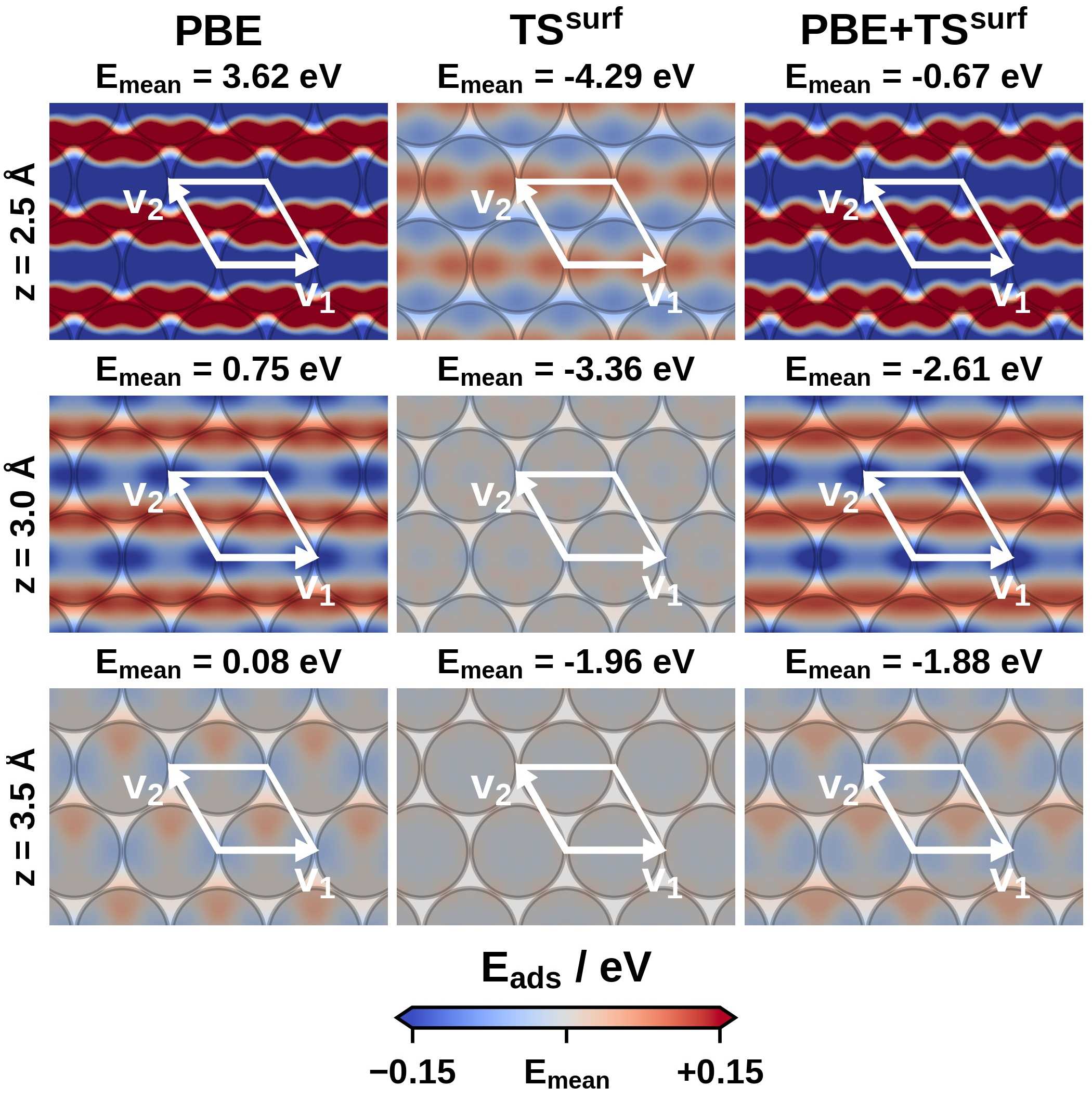}
	\caption{Energy contributions of XC functional and vdW correction to the PES with PTCDA at a constant adsorption height of $2.5~\angstrom$, $3.0~\angstrom$ and $3.5~\angstrom$; The long axis of the molecule is aligned with the primitive substrate lattice vector $\mathbf{v_1}$; The white frame indicates the unit cell}
	\label{fig:corrugation_contribution}
\end{figure}

Indeed, the PESs of PBE with and without TS\textsuperscript{surf} differ by an almost constant energy offset of approximately {\energy{3.36}}, when keeping the PTCDA molecule at a constant height of {\distance{3.00}}. Figure \ref{fig:corrugation_contribution} demonstrates this using the example of PBE+TS\textsuperscript{surf} (section \ref{SIsec:contribution_to_the_adsorption_energy} of the supporting information shows this for all methods).
Hence, the corrugation mainly results from the energy contribution of the functional, rather than from the vdW corrections, and can be tentatively attributed to the laterally varied overlap between orbitals of the molecule and the substrate.
Thus, the vdW correction ``adjusts'' the adsorption height while the XC functional determines the interaction strength and the adsorption sites.


\begin{figure}[H]
	\centering
	\includegraphics[width=\linewidth]{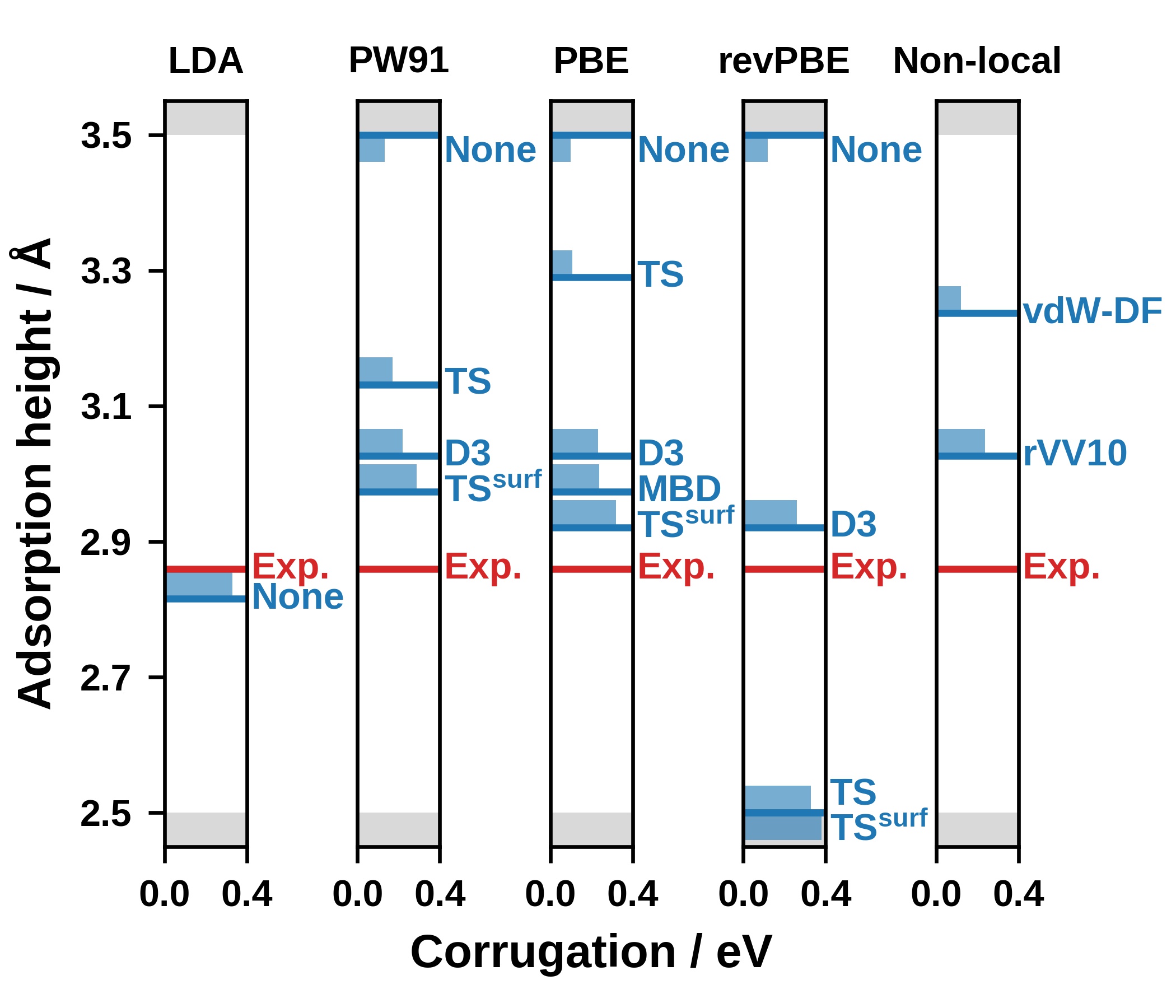}
	\caption{Dependence of the corrugation of the adsorption energy on the adsorption height of the energetic minimum of the molecule aligned with a primitive substrate lattice vector $\mathbf{v_1}$}
	\label{fig:adsorption_height_vs_corrucation}
\end{figure}

Given this insight, we expect that corrugation and adsorption height show an inverse relationship, since a lower adsorption height leads to larger differences in the overlap between states of the molecule and the substrate.
Figure \ref{fig:adsorption_height_vs_corrucation} confirms this expectation. 
It depicts the corrugation of the adsorption energy relative to the optimal adsorption height, with the corrugation being the difference between the energetically highest and lowest point of the two-dimensional cross sections.
In our case, the adsorption height is the distance between the $z$ position of the uppermost substrate layer and the $z$ position of the planar PTCDA molecule.
All GGA functionals show this inverse relationship between corrugation and adsorption height with LDA, vdW-DF and rVV10 following this trend as well (see figure \ref{fig:adsorption_height_vs_corrucation}).
In case of the PW91 and the PBE functionals, the TS\textsuperscript{surf} correction shows the most pronounced corrugation and lowest adsorption height.
In case of revPBE, TS and TS\textsuperscript{surf} yield mean adsorption heights of {\distance{2.50}} (which is the lower boundary of our search space), while showing a larger corrugation than uncorrected revPBE and revPBE+D3.
We will elaborate on the reasons for this behavior in a later part of the results section.

Having taken a look at the relationship between corrugation and adsorption height, we proceed with a quantitative discussion of minima and adsorption geometries.
While previous publications\cite{TSsurf, ruiz2016density} have focused mostly on adsorption heights, our GPR algorithm allows predicting possible adsorption geometries from first principles.
Hence, we not only consider adsorption heights but also investigate molecular orientations and lateral positions of local minima.
To determine these minima, we first use our GPR algorithm to calculate the adsorption energy on a dense grid.
Then we identify all grid points with a minimum value in their respective neighborhood (nearest and next nearest neighbors).
However, some minima may be too shallow and therefore thermodynamically unstable.
Hence, we look for stable local minima, which must be separated from all lower lying minima by an energy barrier of sufficient height.
In our case, a minimum barrier height of {\energy{0.01}} is a good compromise between removing shallow minima and demonstrating the diversity of possible adsorption geometries between the different methods.
We apply this minimum barrier height by using disconnectivity graphs.
These graphs connect all minima via the lowest barrier, allowing for straightforward removal of shallow local minima.
We note that the assumption of a flat rigid molecule and a fixed substrate (see methodology) results in a general overestimation of the energy, i.e., minima may be deeper when accounting for molecular relaxation.
However, since PTCDA is a rigid molecule (see methodology), we assume that this effect is small and has no baring on our interpretation.

\begin{figure}[H]
	\centering
	\includegraphics[width=\linewidth]{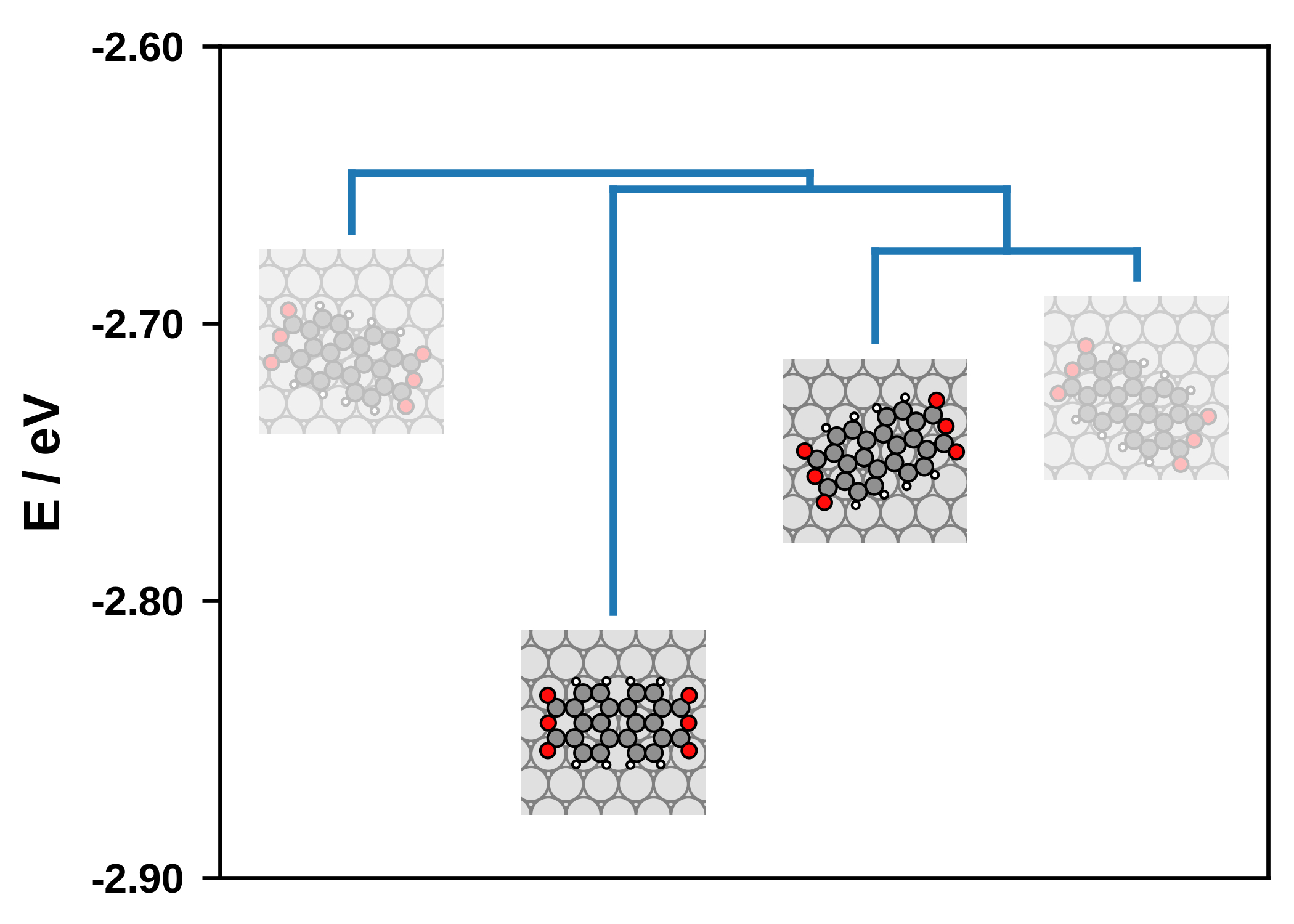}
	\caption{Disconnectivity graph for PBE+TS\textsuperscript{surf}; highlighted adsorption geometries match the experimental adsorption geometries\cite{PhysRevB.76.115421, kraft2006lateral}; \colorize{the lower endpoint of each vertical line denotes the respective binding energy of each minimum and the horizontal lines denote barriers between them}}
	\label{fig:disconnectivity_graph}
\end{figure}

Figure \ref{fig:disconnectivity_graph} shows the disconnectivity graph for PBE+TS\textsuperscript{surf}.
We find four local minima for PBE+TS\textsuperscript{surf}, with the two minima lowest in energy matching the experimentally proposed adsorption geometries.

Disconnectivity graphs for the other combinations of functionals and vdW corrections yield between one and four stable minima (see section \ref{SIsec:disconnectivity_graphs} of the supporting information).
Ideally, the energetically most favorable minima (or minimum if there is only one) correspond to experimental adsorption geometries.
If this is not the case, the list of stable minima should at least contain the experimental geometries.
To test this, we determine the stable minima that are geometrically most similar to the experimental geometries and compare their lateral positions, orientations and adsorption height.
Figure \ref{fig:reproducibility_of_adsorption_geometries} depicts the results, which we will now discuss in detail.

\colorize{LDA yields two stable local minima with adsorption heights of {\distance{2.76}} and {\distance{2.82}} (uncertainty $\Delta h = 0.05~\angstrom$).}
The global minimum is in good agreement with experimental adsorption geometry (B).

All GGA functionals without vdW correction yield minima with adsorption heights of approximately {\distance{3.50}} which is the upper boundary of our search space.
This entails a significant disagreement of more than {\distance{0.60}} between experiment and GGA functionals without vdW correction, and essentially shows that there, the molecule would be described as non-bonded.
Hence, GGA functionals without vdW correction are generally unsuited to calculate PTCDA on Ag(111), which presumably applies equally to other interfaces with strong vdW contributions to their bonding.\cite{Romaner_2009, maurer2019advances}

Augmenting GGA functionals with any vdW correction leads to all methods finding at least one geometry that also occurs in the experiment.
The global minimum of most methods is in good agreement with at least one of the structures found in experiment.
Additionally, adsorption heights determined using D3, TS\textsuperscript{surf} or MBD deviate less than {\distance{0.20}} from experiment.
The exceptions are revPBE+TS and revPBE+TS\textsuperscript{surf}, whose global minima do not correspond to any experimental geometry and display significantly lower adsorption heights than in experiment (see supporting information tables \ref{SItab:adsorption_heights_and_rankings_of_minimum_A} and \ref{SItab:adsorption_heights_and_rankings_of_minimum_B}).

Conversely, PW91+TS\textsuperscript{surf}, PBE+D3 and PBE+TS\textsuperscript{surf} yield both experimental adsorption geometries.
PBE+TS\textsuperscript{surf} conforms best with experiment yielding an adsorption height of {\distance{2.92}}.
This corresponds to a deviation in adsorption height of less than {\distance{0.10}}.
The good performance may in part be due to the fact that PTCDA on Ag(111) was one of the systems used to benchmark the surface parameters for TS\textsuperscript{surf}.\cite{TSsurf}
RevPBE with TS or TS\textsuperscript{surf} also finds both experimental geometries, but places them at adoption heights that are too low --- in three out of four cases at the lower boundary of our search space.
This is due to large attractive contributions of the TS and TS\textsuperscript{surf} correction at low adsorption heights (see section \ref{SIsec:contribution_to_the_adsorption_energy} of the supporting information).
This leads us to assume that the parametrization of TS and TS\textsuperscript{surf} is not adequate when using it with revPBE.
We tentatively attribute this to the $S_R$ parameter, which has been determined using the S22 dataset\cite{jurevcka2006benchmark} and may not be appropriate for interface calculations.
Additionally, in case of revPBE TS/TS\textsuperscript{surf} forces do not improve the {\prediction} accuracy of the GPR algorithm, leading us to exclude them when {\predicting} PESs.

VdW-DF allows finding two stable minima which both match the experimental adsorption geometries.
However, the adsorption heights of {\distance{3.24}} and {\distance{3.29}} are too high, which confirms results from previous studies.\cite{TSsurf, puzder2006binding, toyoda2009first}
rVV10 yields a lower adsorption height of {\distance{2.97}} and finds two minima.
While the global minimum matches experimental geometry (B), the second minimum is slightly offset compared to experimental geometry (A).
This result is comparable those of GGA functionals with vdW corrections, demonstrating that, given our approximations, rVV10 does not offer an improvement over less expensive methods.

All methods, except the GGAs without vdW correction and revPBE with TS or TS\textsuperscript{surf}, find a global minimum matching an experimental adsorption geometry in lateral position and orientation.
Geometry (B) is more readily found than geometry (A).
We find the reason for this behavior by examining the PES of PTCDA aligned with the primitive substrate lattice vector $\mathbf{v_1}$ (alignment with $\mathbf{v_2}$ is symmetry equivalent).
Here, all methods find a PES with trenches along  $\mathbf{v_1}$ --- small differences between methods can easily shift the minimum along the trench and away from the bridge position that corresponds to geometry (A).
In fact all methods with any vdW correction yield a minimum along this trench with the energy difference to the bridge position being smaller than the {\prediction} uncertainty.
Conversely, geometry (B) lies in a potential well surrounded by ridges and is hence easier to find.
Methods yielding both geometries rank geometry (A) lower in energy than geometry (B).
RevPBE with TS or TS\textsuperscript{surf} displays below-average performance.
Both methods yield local minima resembling the experimental geometries, but severely underestimate adsorption heights.
Conversely PW91+TS\textsuperscript{surf}, PBE+D3/TS\textsuperscript{surf} and vdW-DF show good performance with PBE+TS\textsuperscript{surf} coming out on top.

\colorize{It is worth noting that it is by no means a priori clear that both experimentally found geometries must be local minima for individually adsorbed molecules.
Since the herringbone monolayer closely resembles the lateral arrangement in the (102) plane of the $\beta$-PTCDA bulk crystal,\cite{lovinger1984structural, mobus1992structure, ogawa19993} one could also envision a scenario where only one of the geometries is a (sufficiently) deep minimum, and half of the PTCDA molecules assume the second geometry in order to allow for favorable intermolecular interactions.}

\colorize{In general, the structure of close-packed layers results from optimizing the interplay of molecule-molecule and molecule-substrate interactions.\cite{mannsfeld2004analysis, mannsfeld2006advanced}
Nonetheless, the fact that PTCDA on Ag(111) has two local minima that are rotated to each other in such a way that they also allow for very favorable interaction, are almost isoenergetic and have only a very small barrier between them, which can easily be overcome by thermal activation, is an interesting and fortunate coincidence.
It likely contributes to the fact that PTCDA relatively straightforwardly and reproducibly forms commensurate overlayers on Ag(111).
In the context of this work, it is relevant to note that this finding would qualitatively be made with (almost) every method tested here, even if the absolute numbers differ somewhat.}

In summary, we find substantial differences in the optimal adsorption heights, which largely results from differing treatment of vdW interactions.
Notably, GGA functionals without vdW corrections yield adsorption heights at the upper boundary of our search space, wrongly indicating a non-bonded system.

Conversely, lateral features of the PES vary less between different methods.
All methods, except GGAs without vdW corrections, predict a lateral registry of PTCDA on Ag(111) with the caboxylic O atoms seeking positions above Ag atoms.
Hereby, the functionals sufficiently describe the weak covalent bond\cite{hauschild2005molecular} between O and Ag atoms.
Apart from that, the vdW correction is less important for the lateral position and orientation.
This leads to the bridge position being the preferred adsorption site, with few exceptions.

However, there are significant differences in predicted adsorption energies and transition barriers.
PBE+TS\textsuperscript{surf} yields a sufficiently corrugated PES whose minima are separated by large barriers indicating that PTCDA forms commensurate structures on Ag(111).
A similar conclusion could, however, not be drawn as confidently from vdW-DF (to a lesser extent PW91+TS and PBE+D3 as well) where we find a significantly smaller barrier.
Further, commensurability cannot be concluded when using GGAs without vdW corrections since they yield essentially non-bonded systems.

\section{Conclusion}
\label{sc:Conclusion}

We have studied the reproducibility of PESs in relation to the choice of XC functional and vdW correction using the example of PTCDA on Ag(111).
On the one hand, we find a qualitative agreement between combinations of functionals and vdW corrections, on the other hand there exist significant quantitative discrepancies.
Adsorption energies and corrugations vary between different approaches.
The popular PBE functional with the surface dedicated TS\textsuperscript{surf} correction yields the best results.
The two energetically lowest minima of the PES closely resemble the experimental adsorption geometries with adsorption heights differing by less than {\distance{0.10}} between prediction and experiment.
Notably, the energetically most favorable minima of PW91+TS\textsuperscript{surf}, PBE+D3 and vdW-DF also resemble both experimental adsorption geometries, albeit with slightly larger deviations in adsorption height.
Conversely, the GGA functionals fail to capture experimental adsorption geometries when used without vdW corrections.
Hence, describing vdW correctly is essential to get sensible local minima.
The remaining combinations of functionals and vdW corrections yield at least one minimum resembling an experimentally proposed adsorption geometry.
This minimum usually coincides with the global minimum, the only exceptions being revPBE with TS or TS\textsuperscript{surf}.
Notably, separating the PES into contributions of the functional and the vdW correction reveals that the functional determines the lateral position and the position-unspecific vdW adjusts the adsorption height.
Calculating PESs from first principles entails a considerable computational expense requiring the use of 
machine-learning based algorithms such as the one presented here.
PESs often have multiple minima, not all of which are also found experimentally.
Finding the experimental adsorption geometry or the global minimum requires knowledge of the entire PES.

\begin{acknowledgments}
Financial support by the FWF START award (Y 1157-N36) is gratefully acknowledged. Computational results have been achieved using the Vienna Scientific Cluster (VSC).
\end{acknowledgments}

\section*{Data Availability}
\label{sc:data_availability}

\colorize{The data that support the findings of this study are openly available in the NOMAD repository at {\href{https://dx.doi.org/10.17172/NOMAD/2020.08.16-1}{https://dx.doi.org/10.17172/NOMAD/2020.08.16-1}}.\cite{PTCDA_on_Ag111_data}}

\bibliography{Bibliography}

\begin{thebibliography}{72}%
\makeatletter
\providecommand \@ifxundefined [1]{%
 \@ifx{#1\undefined}
}%
\providecommand \@ifnum [1]{%
 \ifnum #1\expandafter \@firstoftwo
 \else \expandafter \@secondoftwo
 \fi
}%
\providecommand \@ifx [1]{%
 \ifx #1\expandafter \@firstoftwo
 \else \expandafter \@secondoftwo
 \fi
}%
\providecommand \natexlab [1]{#1}%
\providecommand \enquote  [1]{``#1''}%
\providecommand \bibnamefont  [1]{#1}%
\providecommand \bibfnamefont [1]{#1}%
\providecommand \citenamefont [1]{#1}%
\providecommand \href@noop [0]{\@secondoftwo}%
\providecommand \href [0]{\begingroup \@sanitize@url \@href}%
\providecommand \@href[1]{\@@startlink{#1}\@@href}%
\providecommand \@@href[1]{\endgroup#1\@@endlink}%
\providecommand \@sanitize@url [0]{\catcode `\\12\catcode `\$12\catcode
  `\&12\catcode `\#12\catcode `\^12\catcode `\_12\catcode `\%12\relax}%
\providecommand \@@startlink[1]{}%
\providecommand \@@endlink[0]{}%
\providecommand \url  [0]{\begingroup\@sanitize@url \@url }%
\providecommand \@url [1]{\endgroup\@href {#1}{\urlprefix }}%
\providecommand \urlprefix  [0]{URL }%
\providecommand \Eprint [0]{\href }%
\providecommand \doibase [0]{http://dx.doi.org/}%
\providecommand \selectlanguage [0]{\@gobble}%
\providecommand \bibinfo  [0]{\@secondoftwo}%
\providecommand \bibfield  [0]{\@secondoftwo}%
\providecommand \translation [1]{[#1]}%
\providecommand \BibitemOpen [0]{}%
\providecommand \bibitemStop [0]{}%
\providecommand \bibitemNoStop [0]{.\EOS\space}%
\providecommand \EOS [0]{\spacefactor3000\relax}%
\providecommand \BibitemShut  [1]{\csname bibitem#1\endcsname}%
\let\auto@bib@innerbib\@empty
\bibitem [{\citenamefont {Tkatchenko}\ \emph {et~al.}(2010)\citenamefont
  {Tkatchenko}, \citenamefont {Romaner}, \citenamefont {Hofmann}, \citenamefont
  {Zojer}, \citenamefont {Ambrosch-Draxl},\ and\ \citenamefont
  {Scheffler}}]{tkatchenko2010van}%
  \BibitemOpen
  \bibfield  {author} {\bibinfo {author} {\bibfnamefont {A.}~\bibnamefont
  {Tkatchenko}}, \bibinfo {author} {\bibfnamefont {L.}~\bibnamefont {Romaner}},
  \bibinfo {author} {\bibfnamefont {O.~T.}\ \bibnamefont {Hofmann}}, \bibinfo
  {author} {\bibfnamefont {E.}~\bibnamefont {Zojer}}, \bibinfo {author}
  {\bibfnamefont {C.}~\bibnamefont {Ambrosch-Draxl}}, \ and\ \bibinfo {author}
  {\bibfnamefont {M.}~\bibnamefont {Scheffler}},\ }\bibfield  {title} {\enquote
  {\bibinfo {title} {Van der waals interactions between organic adsorbates and
  at organic/inorganic interfaces},}\ }\href@noop {} {\bibfield  {journal}
  {\bibinfo  {journal} {MRS bulletin}\ }\textbf {\bibinfo {volume} {35}},\
  \bibinfo {pages} {435--442} (\bibinfo {year} {2010})}\BibitemShut {NoStop}%
\bibitem [{\citenamefont {Rohlfing}, \citenamefont {Temirov},\ and\
  \citenamefont {Tautz}(2007)}]{PhysRevB.76.115421}%
  \BibitemOpen
  \bibfield  {author} {\bibinfo {author} {\bibfnamefont {M.}~\bibnamefont
  {Rohlfing}}, \bibinfo {author} {\bibfnamefont {R.}~\bibnamefont {Temirov}}, \
  and\ \bibinfo {author} {\bibfnamefont {F.~S.}\ \bibnamefont {Tautz}},\
  }\bibfield  {title} {\enquote {\bibinfo {title} {Adsorption structure and
  scanning tunneling data of a prototype organic-inorganic interface: Ptcda on
  ag(111)},}\ }\href {\doibase 10.1103/PhysRevB.76.115421} {\bibfield
  {journal} {\bibinfo  {journal} {Phys. Rev. B}\ }\textbf {\bibinfo {volume}
  {76}},\ \bibinfo {pages} {115421} (\bibinfo {year} {2007})}\BibitemShut
  {NoStop}%
\bibitem [{\citenamefont {Hauschild}\ \emph {et~al.}(2005)\citenamefont
  {Hauschild}, \citenamefont {Karki}, \citenamefont {Cowie}, \citenamefont
  {Rohlfing}, \citenamefont {Tautz},\ and\ \citenamefont
  {Sokolowski}}]{hauschild2005molecular}%
  \BibitemOpen
  \bibfield  {author} {\bibinfo {author} {\bibfnamefont {A.}~\bibnamefont
  {Hauschild}}, \bibinfo {author} {\bibfnamefont {K.}~\bibnamefont {Karki}},
  \bibinfo {author} {\bibfnamefont {B.}~\bibnamefont {Cowie}}, \bibinfo
  {author} {\bibfnamefont {M.}~\bibnamefont {Rohlfing}}, \bibinfo {author}
  {\bibfnamefont {F.}~\bibnamefont {Tautz}}, \ and\ \bibinfo {author}
  {\bibfnamefont {M.}~\bibnamefont {Sokolowski}},\ }\bibfield  {title}
  {\enquote {\bibinfo {title} {Molecular distortions and chemical bonding of a
  large $\pi$-conjugated molecule on a metal surface},}\ }\href@noop {}
  {\bibfield  {journal} {\bibinfo  {journal} {Physical review letters}\
  }\textbf {\bibinfo {volume} {94}},\ \bibinfo {pages} {036106} (\bibinfo
  {year} {2005})}\BibitemShut {NoStop}%
\bibitem [{\citenamefont {H{\"o}rmann}\ \emph {et~al.}(2019)\citenamefont
  {H{\"o}rmann}, \citenamefont {Jeindl}, \citenamefont {Egger}, \citenamefont
  {Scherbela},\ and\ \citenamefont {Hofmann}}]{sample}%
  \BibitemOpen
  \bibfield  {author} {\bibinfo {author} {\bibfnamefont {L.}~\bibnamefont
  {H{\"o}rmann}}, \bibinfo {author} {\bibfnamefont {A.}~\bibnamefont {Jeindl}},
  \bibinfo {author} {\bibfnamefont {A.~T.}\ \bibnamefont {Egger}}, \bibinfo
  {author} {\bibfnamefont {M.}~\bibnamefont {Scherbela}}, \ and\ \bibinfo
  {author} {\bibfnamefont {O.~T.}\ \bibnamefont {Hofmann}},\ }\bibfield
  {title} {\enquote {\bibinfo {title} {Sample: Surface structure search enabled
  by coarse graining and statistical learning},}\ }\href@noop {} {\bibfield
  {journal} {\bibinfo  {journal} {Computer Physics Communications}\ }\textbf
  {\bibinfo {volume} {244}},\ \bibinfo {pages} {143--155} (\bibinfo {year}
  {2019})}\BibitemShut {NoStop}%
\bibitem [{\citenamefont {Gautier}\ \emph {et~al.}(2015)\citenamefont
  {Gautier}, \citenamefont {Steinmann}, \citenamefont {Michel}, \citenamefont
  {Fleurat-Lessard},\ and\ \citenamefont {Sautet}}]{gautier2015molecular}%
  \BibitemOpen
  \bibfield  {author} {\bibinfo {author} {\bibfnamefont {S.}~\bibnamefont
  {Gautier}}, \bibinfo {author} {\bibfnamefont {S.~N.}\ \bibnamefont
  {Steinmann}}, \bibinfo {author} {\bibfnamefont {C.}~\bibnamefont {Michel}},
  \bibinfo {author} {\bibfnamefont {P.}~\bibnamefont {Fleurat-Lessard}}, \ and\
  \bibinfo {author} {\bibfnamefont {P.}~\bibnamefont {Sautet}},\ }\bibfield
  {title} {\enquote {\bibinfo {title} {Molecular adsorption at pt (111). how
  accurate are dft functionals?}}\ }\href@noop {} {\bibfield  {journal}
  {\bibinfo  {journal} {Physical Chemistry Chemical Physics}\ }\textbf
  {\bibinfo {volume} {17}},\ \bibinfo {pages} {28921--28930} (\bibinfo {year}
  {2015})}\BibitemShut {NoStop}%
\bibitem [{\citenamefont {Ehrlich}\ \emph {et~al.}(2011)\citenamefont
  {Ehrlich}, \citenamefont {Moellmann}, \citenamefont {Reckien}, \citenamefont
  {Bredow},\ and\ \citenamefont {Grimme}}]{ehrlich2011system}%
  \BibitemOpen
  \bibfield  {author} {\bibinfo {author} {\bibfnamefont {S.}~\bibnamefont
  {Ehrlich}}, \bibinfo {author} {\bibfnamefont {J.}~\bibnamefont {Moellmann}},
  \bibinfo {author} {\bibfnamefont {W.}~\bibnamefont {Reckien}}, \bibinfo
  {author} {\bibfnamefont {T.}~\bibnamefont {Bredow}}, \ and\ \bibinfo {author}
  {\bibfnamefont {S.}~\bibnamefont {Grimme}},\ }\bibfield  {title} {\enquote
  {\bibinfo {title} {System-dependent dispersion coefficients for the dft-d3
  treatment of adsorption processes on ionic surfaces},}\ }\href@noop {}
  {\bibfield  {journal} {\bibinfo  {journal} {ChemPhysChem}\ }\textbf {\bibinfo
  {volume} {12}},\ \bibinfo {pages} {3414--3420} (\bibinfo {year}
  {2011})}\BibitemShut {NoStop}%
\bibitem [{\citenamefont {Maurer}, \citenamefont {Ruiz},\ and\ \citenamefont
  {Tkatchenko}(2015)}]{maurer2015many}%
  \BibitemOpen
  \bibfield  {author} {\bibinfo {author} {\bibfnamefont {R.~J.}\ \bibnamefont
  {Maurer}}, \bibinfo {author} {\bibfnamefont {V.~G.}\ \bibnamefont {Ruiz}}, \
  and\ \bibinfo {author} {\bibfnamefont {A.}~\bibnamefont {Tkatchenko}},\
  }\bibfield  {title} {\enquote {\bibinfo {title} {Many-body dispersion effects
  in the binding of adsorbates on metal surfaces},}\ }\href@noop {} {\bibfield
  {journal} {\bibinfo  {journal} {The Journal of chemical physics}\ }\textbf
  {\bibinfo {volume} {143}},\ \bibinfo {pages} {102808} (\bibinfo {year}
  {2015})}\BibitemShut {NoStop}%
\bibitem [{\citenamefont {Yang}\ \emph {et~al.}(2010)\citenamefont {Yang},
  \citenamefont {Zheng}, \citenamefont {Zhao},\ and\ \citenamefont
  {Truhlar}}]{yang2010tests}%
  \BibitemOpen
  \bibfield  {author} {\bibinfo {author} {\bibfnamefont {K.}~\bibnamefont
  {Yang}}, \bibinfo {author} {\bibfnamefont {J.}~\bibnamefont {Zheng}},
  \bibinfo {author} {\bibfnamefont {Y.}~\bibnamefont {Zhao}}, \ and\ \bibinfo
  {author} {\bibfnamefont {D.~G.}\ \bibnamefont {Truhlar}},\ }\bibfield
  {title} {\enquote {\bibinfo {title} {Tests of the rpbe, revpbe,
  $\tau$-hcthhyb, $\omega$ b 97 xd, and mohlyp density functional
  approximations and 29 others against representative databases for diverse
  bond energies and barrier heights in catalysis},}\ }\href@noop {} {\bibfield
  {journal} {\bibinfo  {journal} {The Journal of chemical physics}\ }\textbf
  {\bibinfo {volume} {132}},\ \bibinfo {pages} {164117} (\bibinfo {year}
  {2010})}\BibitemShut {NoStop}%
\bibitem [{\citenamefont {Wodrich}\ \emph {et~al.}(2007)\citenamefont
  {Wodrich}, \citenamefont {Corminboeuf}, \citenamefont {Schreiner},
  \citenamefont {Fokin},\ and\ \citenamefont {Schleyer}}]{wodrich2007accurate}%
  \BibitemOpen
  \bibfield  {author} {\bibinfo {author} {\bibfnamefont {M.~D.}\ \bibnamefont
  {Wodrich}}, \bibinfo {author} {\bibfnamefont {C.}~\bibnamefont
  {Corminboeuf}}, \bibinfo {author} {\bibfnamefont {P.~R.}\ \bibnamefont
  {Schreiner}}, \bibinfo {author} {\bibfnamefont {A.~A.}\ \bibnamefont
  {Fokin}}, \ and\ \bibinfo {author} {\bibfnamefont {P.~v.~R.}\ \bibnamefont
  {Schleyer}},\ }\bibfield  {title} {\enquote {\bibinfo {title} {How accurate
  are dft treatments of organic energies?}}\ }\href@noop {} {\bibfield
  {journal} {\bibinfo  {journal} {Organic letters}\ }\textbf {\bibinfo {volume}
  {9}},\ \bibinfo {pages} {1851--1854} (\bibinfo {year} {2007})}\BibitemShut
  {NoStop}%
\bibitem [{\citenamefont {Schultz}, \citenamefont {Zhao},\ and\ \citenamefont
  {Truhlar}(2005)}]{schultz2005databases}%
  \BibitemOpen
  \bibfield  {author} {\bibinfo {author} {\bibfnamefont {N.~E.}\ \bibnamefont
  {Schultz}}, \bibinfo {author} {\bibfnamefont {Y.}~\bibnamefont {Zhao}}, \
  and\ \bibinfo {author} {\bibfnamefont {D.~G.}\ \bibnamefont {Truhlar}},\
  }\bibfield  {title} {\enquote {\bibinfo {title} {Databases for transition
  element bonding: Metal- metal bond energies and bond lengths and their use to
  test hybrid, hybrid meta, and meta density functionals and generalized
  gradient approximations},}\ }\href@noop {} {\bibfield  {journal} {\bibinfo
  {journal} {The Journal of Physical Chemistry A}\ }\textbf {\bibinfo {volume}
  {109}},\ \bibinfo {pages} {4388--4403} (\bibinfo {year} {2005})}\BibitemShut
  {NoStop}%
\bibitem [{\citenamefont {Harvey}(2006)}]{harvey2006accuracy}%
  \BibitemOpen
  \bibfield  {author} {\bibinfo {author} {\bibfnamefont {J.~N.}\ \bibnamefont
  {Harvey}},\ }\bibfield  {title} {\enquote {\bibinfo {title} {On the accuracy
  of density functional theory in transition metal chemistry},}\ }\href@noop {}
  {\bibfield  {journal} {\bibinfo  {journal} {Annual Reports Section"
  C"(Physical Chemistry)}\ }\textbf {\bibinfo {volume} {102}},\ \bibinfo
  {pages} {203--226} (\bibinfo {year} {2006})}\BibitemShut {NoStop}%
\bibitem [{\citenamefont {Zheng}, \citenamefont {Zhao},\ and\ \citenamefont
  {Truhlar}(2009)}]{zheng2009dbh24}%
  \BibitemOpen
  \bibfield  {author} {\bibinfo {author} {\bibfnamefont {J.}~\bibnamefont
  {Zheng}}, \bibinfo {author} {\bibfnamefont {Y.}~\bibnamefont {Zhao}}, \ and\
  \bibinfo {author} {\bibfnamefont {D.~G.}\ \bibnamefont {Truhlar}},\
  }\bibfield  {title} {\enquote {\bibinfo {title} {The dbh24/08 database and
  its use to assess electronic structure model chemistries for chemical
  reaction barrier heights},}\ }\href@noop {} {\bibfield  {journal} {\bibinfo
  {journal} {Journal of Chemical Theory and Computation}\ }\textbf {\bibinfo
  {volume} {5}},\ \bibinfo {pages} {808--821} (\bibinfo {year}
  {2009})}\BibitemShut {NoStop}%
\bibitem [{\citenamefont {Lejaeghere}\ \emph {et~al.}(2016)\citenamefont
  {Lejaeghere}, \citenamefont {Bihlmayer}, \citenamefont {Bj{\"o}rkman},
  \citenamefont {Blaha}, \citenamefont {Bl{\"u}gel}, \citenamefont {Blum},
  \citenamefont {Caliste}, \citenamefont {Castelli}, \citenamefont {Clark},
  \citenamefont {Dal~Corso} \emph {et~al.}}]{lejaeghere2016reproducibility}%
  \BibitemOpen
  \bibfield  {author} {\bibinfo {author} {\bibfnamefont {K.}~\bibnamefont
  {Lejaeghere}}, \bibinfo {author} {\bibfnamefont {G.}~\bibnamefont
  {Bihlmayer}}, \bibinfo {author} {\bibfnamefont {T.}~\bibnamefont
  {Bj{\"o}rkman}}, \bibinfo {author} {\bibfnamefont {P.}~\bibnamefont {Blaha}},
  \bibinfo {author} {\bibfnamefont {S.}~\bibnamefont {Bl{\"u}gel}}, \bibinfo
  {author} {\bibfnamefont {V.}~\bibnamefont {Blum}}, \bibinfo {author}
  {\bibfnamefont {D.}~\bibnamefont {Caliste}}, \bibinfo {author} {\bibfnamefont
  {I.~E.}\ \bibnamefont {Castelli}}, \bibinfo {author} {\bibfnamefont {S.~J.}\
  \bibnamefont {Clark}}, \bibinfo {author} {\bibfnamefont {A.}~\bibnamefont
  {Dal~Corso}},  \emph {et~al.},\ }\bibfield  {title} {\enquote {\bibinfo
  {title} {Reproducibility in density functional theory calculations of
  solids},}\ }\href@noop {} {\bibfield  {journal} {\bibinfo  {journal}
  {Science}\ }\textbf {\bibinfo {volume} {351}},\ \bibinfo {pages} {aad3000}
  (\bibinfo {year} {2016})}\BibitemShut {NoStop}%
\bibitem [{\citenamefont {Ruiz}\ \emph {et~al.}(2012)\citenamefont {Ruiz},
  \citenamefont {Liu}, \citenamefont {Zojer}, \citenamefont {Scheffler},\ and\
  \citenamefont {Tkatchenko}}]{TSsurf}%
  \BibitemOpen
  \bibfield  {author} {\bibinfo {author} {\bibfnamefont {V.~G.}\ \bibnamefont
  {Ruiz}}, \bibinfo {author} {\bibfnamefont {W.}~\bibnamefont {Liu}}, \bibinfo
  {author} {\bibfnamefont {E.}~\bibnamefont {Zojer}}, \bibinfo {author}
  {\bibfnamefont {M.}~\bibnamefont {Scheffler}}, \ and\ \bibinfo {author}
  {\bibfnamefont {A.}~\bibnamefont {Tkatchenko}},\ }\bibfield  {title}
  {\enquote {\bibinfo {title} {Density-functional theory with screened van der
  waals interactions for the modeling of hybrid inorganic-organic systems},}\
  }\href {\doibase 10.1103/PhysRevLett.108.146103} {\bibfield  {journal}
  {\bibinfo  {journal} {Phys. Rev. Lett.}\ }\textbf {\bibinfo {volume} {108}},\
  \bibinfo {pages} {146103} (\bibinfo {year} {2012})}\BibitemShut {NoStop}%
\bibitem [{\citenamefont {Ruiz}, \citenamefont {Liu},\ and\ \citenamefont
  {Tkatchenko}(2016)}]{ruiz2016density}%
  \BibitemOpen
  \bibfield  {author} {\bibinfo {author} {\bibfnamefont {V.~G.}\ \bibnamefont
  {Ruiz}}, \bibinfo {author} {\bibfnamefont {W.}~\bibnamefont {Liu}}, \ and\
  \bibinfo {author} {\bibfnamefont {A.}~\bibnamefont {Tkatchenko}},\ }\bibfield
   {title} {\enquote {\bibinfo {title} {Density-functional theory with screened
  van der waals interactions applied to atomic and molecular adsorbates on
  close-packed and non-close-packed surfaces},}\ }\href@noop {} {\bibfield
  {journal} {\bibinfo  {journal} {Physical Review B}\ }\textbf {\bibinfo
  {volume} {93}},\ \bibinfo {pages} {035118} (\bibinfo {year}
  {2016})}\BibitemShut {NoStop}%
\bibitem [{\citenamefont {Liu}\ \emph {et~al.}(2015)\citenamefont {Liu},
  \citenamefont {Maa{\ss}}, \citenamefont {Willenbockel}, \citenamefont
  {Bronner}, \citenamefont {Schulze}, \citenamefont {Soubatch}, \citenamefont
  {Tautz}, \citenamefont {Tegeder},\ and\ \citenamefont
  {Tkatchenko}}]{liu2015quantitative}%
  \BibitemOpen
  \bibfield  {author} {\bibinfo {author} {\bibfnamefont {W.}~\bibnamefont
  {Liu}}, \bibinfo {author} {\bibfnamefont {F.}~\bibnamefont {Maa{\ss}}},
  \bibinfo {author} {\bibfnamefont {M.}~\bibnamefont {Willenbockel}}, \bibinfo
  {author} {\bibfnamefont {C.}~\bibnamefont {Bronner}}, \bibinfo {author}
  {\bibfnamefont {M.}~\bibnamefont {Schulze}}, \bibinfo {author} {\bibfnamefont
  {S.}~\bibnamefont {Soubatch}}, \bibinfo {author} {\bibfnamefont {F.~S.}\
  \bibnamefont {Tautz}}, \bibinfo {author} {\bibfnamefont {P.}~\bibnamefont
  {Tegeder}}, \ and\ \bibinfo {author} {\bibfnamefont {A.}~\bibnamefont
  {Tkatchenko}},\ }\bibfield  {title} {\enquote {\bibinfo {title} {Quantitative
  prediction of molecular adsorption: structure and binding of benzene on
  coinage metals},}\ }\href@noop {} {\bibfield  {journal} {\bibinfo  {journal}
  {Physical review letters}\ }\textbf {\bibinfo {volume} {115}},\ \bibinfo
  {pages} {036104} (\bibinfo {year} {2015})}\BibitemShut {NoStop}%
\bibitem [{\citenamefont {Reckien}, \citenamefont {Eggers},\ and\ \citenamefont
  {Bredow}(2014)}]{reckien2014theoretical}%
  \BibitemOpen
  \bibfield  {author} {\bibinfo {author} {\bibfnamefont {W.}~\bibnamefont
  {Reckien}}, \bibinfo {author} {\bibfnamefont {M.}~\bibnamefont {Eggers}}, \
  and\ \bibinfo {author} {\bibfnamefont {T.}~\bibnamefont {Bredow}},\
  }\bibfield  {title} {\enquote {\bibinfo {title} {Theoretical study of the
  adsorption of benzene on coinage metals},}\ }\href@noop {} {\bibfield
  {journal} {\bibinfo  {journal} {Beilstein journal of organic chemistry}\
  }\textbf {\bibinfo {volume} {10}},\ \bibinfo {pages} {1775--1784} (\bibinfo
  {year} {2014})}\BibitemShut {NoStop}%
\bibitem [{\citenamefont {Liu}\ \emph {et~al.}(2013)\citenamefont {Liu},
  \citenamefont {Ruiz}, \citenamefont {Zhang}, \citenamefont {Santra},
  \citenamefont {Ren}, \citenamefont {Scheffler},\ and\ \citenamefont
  {Tkatchenko}}]{liu2013structure}%
  \BibitemOpen
  \bibfield  {author} {\bibinfo {author} {\bibfnamefont {W.}~\bibnamefont
  {Liu}}, \bibinfo {author} {\bibfnamefont {V.~G.}\ \bibnamefont {Ruiz}},
  \bibinfo {author} {\bibfnamefont {G.-X.}\ \bibnamefont {Zhang}}, \bibinfo
  {author} {\bibfnamefont {B.}~\bibnamefont {Santra}}, \bibinfo {author}
  {\bibfnamefont {X.}~\bibnamefont {Ren}}, \bibinfo {author} {\bibfnamefont
  {M.}~\bibnamefont {Scheffler}}, \ and\ \bibinfo {author} {\bibfnamefont
  {A.}~\bibnamefont {Tkatchenko}},\ }\bibfield  {title} {\enquote {\bibinfo
  {title} {Structure and energetics of benzene adsorbed on transition-metal
  surfaces: density-functional theory with van der waals interactions including
  collective substrate response},}\ }\href@noop {} {\bibfield  {journal}
  {\bibinfo  {journal} {New Journal of Physics}\ }\textbf {\bibinfo {volume}
  {15}},\ \bibinfo {pages} {053046} (\bibinfo {year} {2013})}\BibitemShut
  {NoStop}%
\bibitem [{\citenamefont {Li}\ \emph {et~al.}(2012)\citenamefont {Li},
  \citenamefont {Tamblyn}, \citenamefont {Cooper}, \citenamefont {Gao},\ and\
  \citenamefont {Neaton}}]{li2012molecular}%
  \BibitemOpen
  \bibfield  {author} {\bibinfo {author} {\bibfnamefont {G.}~\bibnamefont
  {Li}}, \bibinfo {author} {\bibfnamefont {I.}~\bibnamefont {Tamblyn}},
  \bibinfo {author} {\bibfnamefont {V.~R.}\ \bibnamefont {Cooper}}, \bibinfo
  {author} {\bibfnamefont {H.-J.}\ \bibnamefont {Gao}}, \ and\ \bibinfo
  {author} {\bibfnamefont {J.~B.}\ \bibnamefont {Neaton}},\ }\bibfield  {title}
  {\enquote {\bibinfo {title} {Molecular adsorption on metal surfaces with van
  der waals density functionals},}\ }\href@noop {} {\bibfield  {journal}
  {\bibinfo  {journal} {Physical Review B}\ }\textbf {\bibinfo {volume} {85}},\
  \bibinfo {pages} {121409} (\bibinfo {year} {2012})}\BibitemShut {NoStop}%
\bibitem [{\citenamefont {Romaner}\ \emph {et~al.}(2009)\citenamefont
  {Romaner}, \citenamefont {Nabok}, \citenamefont {Puschnig}, \citenamefont
  {Zojer},\ and\ \citenamefont {Ambrosch-Draxl}}]{Romaner_2009}%
  \BibitemOpen
  \bibfield  {author} {\bibinfo {author} {\bibfnamefont {L.}~\bibnamefont
  {Romaner}}, \bibinfo {author} {\bibfnamefont {D.}~\bibnamefont {Nabok}},
  \bibinfo {author} {\bibfnamefont {P.}~\bibnamefont {Puschnig}}, \bibinfo
  {author} {\bibfnamefont {E.}~\bibnamefont {Zojer}}, \ and\ \bibinfo {author}
  {\bibfnamefont {C.}~\bibnamefont {Ambrosch-Draxl}},\ }\bibfield  {title}
  {\enquote {\bibinfo {title} {Theoretical study of {PTCDA} adsorbed on the
  coinage metal surfaces, ag(111), au(111) and cu(111)},}\ }\href {\doibase
  10.1088/1367-2630/11/5/053010} {\bibfield  {journal} {\bibinfo  {journal}
  {New Journal of Physics}\ }\textbf {\bibinfo {volume} {11}},\ \bibinfo
  {pages} {053010} (\bibinfo {year} {2009})}\BibitemShut {NoStop}%
\bibitem [{\citenamefont {Kraft}\ \emph {et~al.}(2006)\citenamefont {Kraft},
  \citenamefont {Temirov}, \citenamefont {Henze}, \citenamefont {Soubatch},
  \citenamefont {Rohlfing},\ and\ \citenamefont {Tautz}}]{kraft2006lateral}%
  \BibitemOpen
  \bibfield  {author} {\bibinfo {author} {\bibfnamefont {A.}~\bibnamefont
  {Kraft}}, \bibinfo {author} {\bibfnamefont {R.}~\bibnamefont {Temirov}},
  \bibinfo {author} {\bibfnamefont {S.}~\bibnamefont {Henze}}, \bibinfo
  {author} {\bibfnamefont {S.}~\bibnamefont {Soubatch}}, \bibinfo {author}
  {\bibfnamefont {M.}~\bibnamefont {Rohlfing}}, \ and\ \bibinfo {author}
  {\bibfnamefont {F.}~\bibnamefont {Tautz}},\ }\bibfield  {title} {\enquote
  {\bibinfo {title} {Lateral adsorption geometry and site-specific electronic
  structure of a large organic chemisorbate on a metal surface},}\ }\href@noop
  {} {\bibfield  {journal} {\bibinfo  {journal} {Physical Review B}\ }\textbf
  {\bibinfo {volume} {74}},\ \bibinfo {pages} {041402} (\bibinfo {year}
  {2006})}\BibitemShut {NoStop}%
\bibitem [{\citenamefont {Hooks}, \citenamefont {Fritz},\ and\ \citenamefont
  {Ward}(2001)}]{hooks2001epitaxy}%
  \BibitemOpen
  \bibfield  {author} {\bibinfo {author} {\bibfnamefont {D.~E.}\ \bibnamefont
  {Hooks}}, \bibinfo {author} {\bibfnamefont {T.}~\bibnamefont {Fritz}}, \ and\
  \bibinfo {author} {\bibfnamefont {M.~D.}\ \bibnamefont {Ward}},\ }\bibfield
  {title} {\enquote {\bibinfo {title} {Epitaxy and molecular organization on
  solid substrates},}\ }\href@noop {} {\bibfield  {journal} {\bibinfo
  {journal} {Advanced Materials}\ }\textbf {\bibinfo {volume} {13}},\ \bibinfo
  {pages} {227--241} (\bibinfo {year} {2001})}\BibitemShut {NoStop}%
\bibitem [{\citenamefont {Ceperley}\ and\ \citenamefont {Alder}(1980)}]{lda}%
  \BibitemOpen
  \bibfield  {author} {\bibinfo {author} {\bibfnamefont {D.~M.}\ \bibnamefont
  {Ceperley}}\ and\ \bibinfo {author} {\bibfnamefont {B.~J.}\ \bibnamefont
  {Alder}},\ }\bibfield  {title} {\enquote {\bibinfo {title} {Ground state of
  the electron gas by a stochastic method},}\ }\href {\doibase
  10.1103/PhysRevLett.45.566} {\bibfield  {journal} {\bibinfo  {journal} {Phys.
  Rev. Lett.}\ }\textbf {\bibinfo {volume} {45}},\ \bibinfo {pages} {566--569}
  (\bibinfo {year} {1980})}\BibitemShut {NoStop}%
\bibitem [{\citenamefont {Lee}\ and\ \citenamefont {Yu}(2005)}]{lee2005ab}%
  \BibitemOpen
  \bibfield  {author} {\bibinfo {author} {\bibfnamefont {K.}~\bibnamefont
  {Lee}}\ and\ \bibinfo {author} {\bibfnamefont {J.}~\bibnamefont {Yu}},\
  }\bibfield  {title} {\enquote {\bibinfo {title} {Ab initio study of pentacene
  on au (0 0 1) surface},}\ }\href@noop {} {\bibfield  {journal} {\bibinfo
  {journal} {Surface science}\ }\textbf {\bibinfo {volume} {589}},\ \bibinfo
  {pages} {8--18} (\bibinfo {year} {2005})}\BibitemShut {NoStop}%
\bibitem [{\citenamefont {Tournus}\ and\ \citenamefont
  {Charlier}(2005)}]{tournus2005ab}%
  \BibitemOpen
  \bibfield  {author} {\bibinfo {author} {\bibfnamefont {F.}~\bibnamefont
  {Tournus}}\ and\ \bibinfo {author} {\bibfnamefont {J.-C.}\ \bibnamefont
  {Charlier}},\ }\bibfield  {title} {\enquote {\bibinfo {title} {Ab initio
  study of benzene adsorption on carbon nanotubes},}\ }\href@noop {} {\bibfield
   {journal} {\bibinfo  {journal} {Physical Review B}\ }\textbf {\bibinfo
  {volume} {71}},\ \bibinfo {pages} {165421} (\bibinfo {year}
  {2005})}\BibitemShut {NoStop}%
\bibitem [{\citenamefont {Perdew}\ \emph {et~al.}(1992)\citenamefont {Perdew},
  \citenamefont {Chevary}, \citenamefont {Vosko}, \citenamefont {Jackson},
  \citenamefont {Pederson}, \citenamefont {Singh},\ and\ \citenamefont
  {Fiolhais}}]{pw91}%
  \BibitemOpen
  \bibfield  {author} {\bibinfo {author} {\bibfnamefont {J.~P.}\ \bibnamefont
  {Perdew}}, \bibinfo {author} {\bibfnamefont {J.~A.}\ \bibnamefont {Chevary}},
  \bibinfo {author} {\bibfnamefont {S.~H.}\ \bibnamefont {Vosko}}, \bibinfo
  {author} {\bibfnamefont {K.~A.}\ \bibnamefont {Jackson}}, \bibinfo {author}
  {\bibfnamefont {M.~R.}\ \bibnamefont {Pederson}}, \bibinfo {author}
  {\bibfnamefont {D.~J.}\ \bibnamefont {Singh}}, \ and\ \bibinfo {author}
  {\bibfnamefont {C.}~\bibnamefont {Fiolhais}},\ }\bibfield  {title} {\enquote
  {\bibinfo {title} {Atoms, molecules, solids, and surfaces: Applications of
  the generalized gradient approximation for exchange and correlation},}\
  }\href {\doibase 10.1103/PhysRevB.46.6671} {\bibfield  {journal} {\bibinfo
  {journal} {Phys. Rev. B}\ }\textbf {\bibinfo {volume} {46}},\ \bibinfo
  {pages} {6671--6687} (\bibinfo {year} {1992})}\BibitemShut {NoStop}%
\bibitem [{\citenamefont {Perdew}, \citenamefont {Burke},\ and\ \citenamefont
  {Ernzerhof}(1996)}]{pbe}%
  \BibitemOpen
  \bibfield  {author} {\bibinfo {author} {\bibfnamefont {J.~P.}\ \bibnamefont
  {Perdew}}, \bibinfo {author} {\bibfnamefont {K.}~\bibnamefont {Burke}}, \
  and\ \bibinfo {author} {\bibfnamefont {M.}~\bibnamefont {Ernzerhof}},\
  }\bibfield  {title} {\enquote {\bibinfo {title} {Generalized gradient
  approximation made simple},}\ }\href {\doibase 10.1103/PhysRevLett.77.3865}
  {\bibfield  {journal} {\bibinfo  {journal} {Phys. Rev. Lett.}\ }\textbf
  {\bibinfo {volume} {77}},\ \bibinfo {pages} {3865--3868} (\bibinfo {year}
  {1996})}\BibitemShut {NoStop}%
\bibitem [{\citenamefont {Zhang}\ and\ \citenamefont {Yang}(1998)}]{revpbe}%
  \BibitemOpen
  \bibfield  {author} {\bibinfo {author} {\bibfnamefont {Y.}~\bibnamefont
  {Zhang}}\ and\ \bibinfo {author} {\bibfnamefont {W.}~\bibnamefont {Yang}},\
  }\bibfield  {title} {\enquote {\bibinfo {title} {Comment on ``generalized
  gradient approximation made simple''},}\ }\href {\doibase
  10.1103/PhysRevLett.80.890} {\bibfield  {journal} {\bibinfo  {journal} {Phys.
  Rev. Lett.}\ }\textbf {\bibinfo {volume} {80}},\ \bibinfo {pages} {890--890}
  (\bibinfo {year} {1998})}\BibitemShut {NoStop}%
\bibitem [{\citenamefont {Burke}(2012)}]{burke2012perspective}%
  \BibitemOpen
  \bibfield  {author} {\bibinfo {author} {\bibfnamefont {K.}~\bibnamefont
  {Burke}},\ }\bibfield  {title} {\enquote {\bibinfo {title} {Perspective on
  density functional theory},}\ }\href@noop {} {\bibfield  {journal} {\bibinfo
  {journal} {The Journal of chemical physics}\ }\textbf {\bibinfo {volume}
  {136}},\ \bibinfo {pages} {150901} (\bibinfo {year} {2012})}\BibitemShut
  {NoStop}%
\bibitem [{\citenamefont {Maurer}\ \emph {et~al.}(2019)\citenamefont {Maurer},
  \citenamefont {Freysoldt}, \citenamefont {Reilly}, \citenamefont
  {Brandenburg}, \citenamefont {Hofmann}, \citenamefont {Bj{\"o}rkman},
  \citenamefont {Leb{\`e}gue},\ and\ \citenamefont
  {Tkatchenko}}]{maurer2019advances}%
  \BibitemOpen
  \bibfield  {author} {\bibinfo {author} {\bibfnamefont {R.~J.}\ \bibnamefont
  {Maurer}}, \bibinfo {author} {\bibfnamefont {C.}~\bibnamefont {Freysoldt}},
  \bibinfo {author} {\bibfnamefont {A.~M.}\ \bibnamefont {Reilly}}, \bibinfo
  {author} {\bibfnamefont {J.~G.}\ \bibnamefont {Brandenburg}}, \bibinfo
  {author} {\bibfnamefont {O.~T.}\ \bibnamefont {Hofmann}}, \bibinfo {author}
  {\bibfnamefont {T.}~\bibnamefont {Bj{\"o}rkman}}, \bibinfo {author}
  {\bibfnamefont {S.}~\bibnamefont {Leb{\`e}gue}}, \ and\ \bibinfo {author}
  {\bibfnamefont {A.}~\bibnamefont {Tkatchenko}},\ }\bibfield  {title}
  {\enquote {\bibinfo {title} {Advances in density-functional calculations for
  materials modeling},}\ }\href@noop {} {\bibfield  {journal} {\bibinfo
  {journal} {Annual Review of Materials Research}\ }\textbf {\bibinfo {volume}
  {49}},\ \bibinfo {pages} {1--30} (\bibinfo {year} {2019})}\BibitemShut
  {NoStop}%
\bibitem [{\citenamefont {Hammer}, \citenamefont {Hansen},\ and\ \citenamefont
  {N{\o}rskov}(1999)}]{hammer1999improved}%
  \BibitemOpen
  \bibfield  {author} {\bibinfo {author} {\bibfnamefont {B.}~\bibnamefont
  {Hammer}}, \bibinfo {author} {\bibfnamefont {L.~B.}\ \bibnamefont {Hansen}},
  \ and\ \bibinfo {author} {\bibfnamefont {J.~K.}\ \bibnamefont {N{\o}rskov}},\
  }\bibfield  {title} {\enquote {\bibinfo {title} {Improved adsorption
  energetics within density-functional theory using revised
  perdew-burke-ernzerhof functionals},}\ }\href@noop {} {\bibfield  {journal}
  {\bibinfo  {journal} {Physical review B}\ }\textbf {\bibinfo {volume} {59}},\
  \bibinfo {pages} {7413} (\bibinfo {year} {1999})}\BibitemShut {NoStop}%
\bibitem [{\citenamefont {Grimme}\ \emph {et~al.}(2010)\citenamefont {Grimme},
  \citenamefont {Antony}, \citenamefont {Ehrlich},\ and\ \citenamefont
  {Krieg}}]{D3}%
  \BibitemOpen
  \bibfield  {author} {\bibinfo {author} {\bibfnamefont {S.}~\bibnamefont
  {Grimme}}, \bibinfo {author} {\bibfnamefont {J.}~\bibnamefont {Antony}},
  \bibinfo {author} {\bibfnamefont {S.}~\bibnamefont {Ehrlich}}, \ and\
  \bibinfo {author} {\bibfnamefont {H.}~\bibnamefont {Krieg}},\ }\bibfield
  {title} {\enquote {\bibinfo {title} {A consistent and accurate ab initio
  parametrization of density functional dispersion correction (dft-d) for the
  94 elements h-pu},}\ }\href@noop {} {\bibfield  {journal} {\bibinfo
  {journal} {The Journal of chemical physics}\ }\textbf {\bibinfo {volume}
  {132}},\ \bibinfo {pages} {154104} (\bibinfo {year} {2010})}\BibitemShut
  {NoStop}%
\bibitem [{\citenamefont {Grimme}, \citenamefont {Ehrlich},\ and\ \citenamefont
  {Goerigk}(2011)}]{grimme2011effect}%
  \BibitemOpen
  \bibfield  {author} {\bibinfo {author} {\bibfnamefont {S.}~\bibnamefont
  {Grimme}}, \bibinfo {author} {\bibfnamefont {S.}~\bibnamefont {Ehrlich}}, \
  and\ \bibinfo {author} {\bibfnamefont {L.}~\bibnamefont {Goerigk}},\
  }\bibfield  {title} {\enquote {\bibinfo {title} {Effect of the damping
  function in dispersion corrected density functional theory},}\ }\href@noop {}
  {\bibfield  {journal} {\bibinfo  {journal} {Journal of computational
  chemistry}\ }\textbf {\bibinfo {volume} {32}},\ \bibinfo {pages} {1456--1465}
  (\bibinfo {year} {2011})}\BibitemShut {NoStop}%
\bibitem [{\citenamefont {Grimme}(2006)}]{TS}%
  \BibitemOpen
  \bibfield  {author} {\bibinfo {author} {\bibfnamefont {S.}~\bibnamefont
  {Grimme}},\ }\bibfield  {title} {\enquote {\bibinfo {title} {Semiempirical
  gga-type density functional constructed with a long-range dispersion
  correction},}\ }\href@noop {} {\bibfield  {journal} {\bibinfo  {journal}
  {Journal of computational chemistry}\ }\textbf {\bibinfo {volume} {27}},\
  \bibinfo {pages} {1787--1799} (\bibinfo {year} {2006})}\BibitemShut {NoStop}%
\bibitem [{\citenamefont {Tkatchenko}\ \emph {et~al.}(2012)\citenamefont
  {Tkatchenko}, \citenamefont {DiStasio}, \citenamefont {Car},\ and\
  \citenamefont {Scheffler}}]{MBD}%
  \BibitemOpen
  \bibfield  {author} {\bibinfo {author} {\bibfnamefont {A.}~\bibnamefont
  {Tkatchenko}}, \bibinfo {author} {\bibfnamefont {R.~A.}\ \bibnamefont
  {DiStasio}}, \bibinfo {author} {\bibfnamefont {R.}~\bibnamefont {Car}}, \
  and\ \bibinfo {author} {\bibfnamefont {M.}~\bibnamefont {Scheffler}},\
  }\bibfield  {title} {\enquote {\bibinfo {title} {Accurate and efficient
  method for many-body van der waals interactions},}\ }\href {\doibase
  10.1103/PhysRevLett.108.236402} {\bibfield  {journal} {\bibinfo  {journal}
  {Phys. Rev. Lett.}\ }\textbf {\bibinfo {volume} {108}},\ \bibinfo {pages}
  {236402} (\bibinfo {year} {2012})}\BibitemShut {NoStop}%
\bibitem [{\citenamefont {Blood-Forsythe}\ \emph {et~al.}(2016)\citenamefont
  {Blood-Forsythe}, \citenamefont {Markovich}, \citenamefont {DiStasio},
  \citenamefont {Car},\ and\ \citenamefont
  {Aspuru-Guzik}}]{blood2016analytical}%
  \BibitemOpen
  \bibfield  {author} {\bibinfo {author} {\bibfnamefont {M.~A.}\ \bibnamefont
  {Blood-Forsythe}}, \bibinfo {author} {\bibfnamefont {T.}~\bibnamefont
  {Markovich}}, \bibinfo {author} {\bibfnamefont {R.~A.}\ \bibnamefont
  {DiStasio}}, \bibinfo {author} {\bibfnamefont {R.}~\bibnamefont {Car}}, \
  and\ \bibinfo {author} {\bibfnamefont {A.}~\bibnamefont {Aspuru-Guzik}},\
  }\bibfield  {title} {\enquote {\bibinfo {title} {Analytical nuclear gradients
  for the range-separated many-body dispersion model of noncovalent
  interactions},}\ }\href@noop {} {\bibfield  {journal} {\bibinfo  {journal}
  {Chemical science}\ }\textbf {\bibinfo {volume} {7}},\ \bibinfo {pages}
  {1712--1728} (\bibinfo {year} {2016})}\BibitemShut {NoStop}%
\bibitem [{\citenamefont {Dion}\ \emph {et~al.}(2004)\citenamefont {Dion},
  \citenamefont {Rydberg}, \citenamefont {Schr\"oder}, \citenamefont
  {Langreth},\ and\ \citenamefont {Lundqvist}}]{vdwDF}%
  \BibitemOpen
  \bibfield  {author} {\bibinfo {author} {\bibfnamefont {M.}~\bibnamefont
  {Dion}}, \bibinfo {author} {\bibfnamefont {H.}~\bibnamefont {Rydberg}},
  \bibinfo {author} {\bibfnamefont {E.}~\bibnamefont {Schr\"oder}}, \bibinfo
  {author} {\bibfnamefont {D.~C.}\ \bibnamefont {Langreth}}, \ and\ \bibinfo
  {author} {\bibfnamefont {B.~I.}\ \bibnamefont {Lundqvist}},\ }\bibfield
  {title} {\enquote {\bibinfo {title} {Van der waals density functional for
  general geometries},}\ }\href {\doibase 10.1103/PhysRevLett.92.246401}
  {\bibfield  {journal} {\bibinfo  {journal} {Phys. Rev. Lett.}\ }\textbf
  {\bibinfo {volume} {92}},\ \bibinfo {pages} {246401} (\bibinfo {year}
  {2004})}\BibitemShut {NoStop}%
\bibitem [{\citenamefont {Xu}\ \emph {et~al.}(2013)\citenamefont {Xu},
  \citenamefont {Hofmann}, \citenamefont {Schlesinger}, \citenamefont
  {Winkler}, \citenamefont {Frisch}, \citenamefont {Niederhausen},
  \citenamefont {Vollmer}, \citenamefont {Blumstengel}, \citenamefont
  {Henneberger}, \citenamefont {Koch} \emph {et~al.}}]{xu2013space}%
  \BibitemOpen
  \bibfield  {author} {\bibinfo {author} {\bibfnamefont {Y.}~\bibnamefont
  {Xu}}, \bibinfo {author} {\bibfnamefont {O.~T.}\ \bibnamefont {Hofmann}},
  \bibinfo {author} {\bibfnamefont {R.}~\bibnamefont {Schlesinger}}, \bibinfo
  {author} {\bibfnamefont {S.}~\bibnamefont {Winkler}}, \bibinfo {author}
  {\bibfnamefont {J.}~\bibnamefont {Frisch}}, \bibinfo {author} {\bibfnamefont
  {J.}~\bibnamefont {Niederhausen}}, \bibinfo {author} {\bibfnamefont
  {A.}~\bibnamefont {Vollmer}}, \bibinfo {author} {\bibfnamefont
  {S.}~\bibnamefont {Blumstengel}}, \bibinfo {author} {\bibfnamefont
  {F.}~\bibnamefont {Henneberger}}, \bibinfo {author} {\bibfnamefont
  {N.}~\bibnamefont {Koch}},  \emph {et~al.},\ }\bibfield  {title} {\enquote
  {\bibinfo {title} {Space-charge transfer in hybrid inorganic-organic
  systems},}\ }\href@noop {} {\bibfield  {journal} {\bibinfo  {journal}
  {Physical review letters}\ }\textbf {\bibinfo {volume} {111}},\ \bibinfo
  {pages} {226802} (\bibinfo {year} {2013})}\BibitemShut {NoStop}%
\bibitem [{\citenamefont {Wang}\ \emph {et~al.}(2019)\citenamefont {Wang},
  \citenamefont {Levchenko}, \citenamefont {Schultz}, \citenamefont {Koch},
  \citenamefont {Scheffler},\ and\ \citenamefont {Rossi}}]{wang2019modulation}%
  \BibitemOpen
  \bibfield  {author} {\bibinfo {author} {\bibfnamefont {H.}~\bibnamefont
  {Wang}}, \bibinfo {author} {\bibfnamefont {S.~V.}\ \bibnamefont {Levchenko}},
  \bibinfo {author} {\bibfnamefont {T.}~\bibnamefont {Schultz}}, \bibinfo
  {author} {\bibfnamefont {N.}~\bibnamefont {Koch}}, \bibinfo {author}
  {\bibfnamefont {M.}~\bibnamefont {Scheffler}}, \ and\ \bibinfo {author}
  {\bibfnamefont {M.}~\bibnamefont {Rossi}},\ }\bibfield  {title} {\enquote
  {\bibinfo {title} {Modulation of the work function by the atomic structure of
  strong organic electron acceptors on h-si (111)},}\ }\href@noop {} {\bibfield
   {journal} {\bibinfo  {journal} {Advanced Electronic Materials}\ }\textbf
  {\bibinfo {volume} {5}},\ \bibinfo {pages} {1800891} (\bibinfo {year}
  {2019})}\BibitemShut {NoStop}%
\bibitem [{\citenamefont {Hofmann}\ \emph {et~al.}(2015)\citenamefont
  {Hofmann}, \citenamefont {Rinke}, \citenamefont {Scheffler},\ and\
  \citenamefont {Heimel}}]{hofmann2015integer}%
  \BibitemOpen
  \bibfield  {author} {\bibinfo {author} {\bibfnamefont {O.~T.}\ \bibnamefont
  {Hofmann}}, \bibinfo {author} {\bibfnamefont {P.}~\bibnamefont {Rinke}},
  \bibinfo {author} {\bibfnamefont {M.}~\bibnamefont {Scheffler}}, \ and\
  \bibinfo {author} {\bibfnamefont {G.}~\bibnamefont {Heimel}},\ }\bibfield
  {title} {\enquote {\bibinfo {title} {Integer versus fractional charge
  transfer at metal (/insulator)/organic interfaces: Cu (/nacl)/tcne},}\
  }\href@noop {} {\bibfield  {journal} {\bibinfo  {journal} {ACS nano}\
  }\textbf {\bibinfo {volume} {9}},\ \bibinfo {pages} {5391--5404} (\bibinfo
  {year} {2015})}\BibitemShut {NoStop}%
\bibitem [{\citenamefont {Gruenewald}\ \emph {et~al.}(2015)\citenamefont
  {Gruenewald}, \citenamefont {Schirra}, \citenamefont {Winget}, \citenamefont
  {Kozlik}, \citenamefont {Ndione}, \citenamefont {Sigdel}, \citenamefont
  {Berry}, \citenamefont {Forker}, \citenamefont {Brédas}, \citenamefont
  {Fritz},\ and\ \citenamefont {Monti}}]{gruenewald2015integer}%
  \BibitemOpen
  \bibfield  {author} {\bibinfo {author} {\bibfnamefont {M.}~\bibnamefont
  {Gruenewald}}, \bibinfo {author} {\bibfnamefont {L.~K.}\ \bibnamefont
  {Schirra}}, \bibinfo {author} {\bibfnamefont {P.}~\bibnamefont {Winget}},
  \bibinfo {author} {\bibfnamefont {M.}~\bibnamefont {Kozlik}}, \bibinfo
  {author} {\bibfnamefont {P.~F.}\ \bibnamefont {Ndione}}, \bibinfo {author}
  {\bibfnamefont {A.~K.}\ \bibnamefont {Sigdel}}, \bibinfo {author}
  {\bibfnamefont {J.~J.}\ \bibnamefont {Berry}}, \bibinfo {author}
  {\bibfnamefont {R.}~\bibnamefont {Forker}}, \bibinfo {author} {\bibfnamefont
  {J.-L.}\ \bibnamefont {Brédas}}, \bibinfo {author} {\bibfnamefont
  {T.}~\bibnamefont {Fritz}}, \ and\ \bibinfo {author} {\bibfnamefont
  {O.~L.~A.}\ \bibnamefont {Monti}},\ }\bibfield  {title} {\enquote {\bibinfo
  {title} {Integer charge transfer and hybridization at an organic
  semiconductor/conductive oxide interface},}\ }\href {\doibase
  10.1021/jp512153b} {\bibfield  {journal} {\bibinfo  {journal} {The Journal of
  Physical Chemistry C}\ }\textbf {\bibinfo {volume} {119}},\ \bibinfo {pages}
  {4865--4873} (\bibinfo {year} {2015})},\ \Eprint
  {http://arxiv.org/abs/https://doi.org/10.1021/jp512153b}
  {https://doi.org/10.1021/jp512153b} \BibitemShut {NoStop}%
\bibitem [{\citenamefont {Moll}\ \emph {et~al.}(2013)\citenamefont {Moll},
  \citenamefont {Xu}, \citenamefont {Hofmann},\ and\ \citenamefont
  {Rinke}}]{moll2013stabilization}%
  \BibitemOpen
  \bibfield  {author} {\bibinfo {author} {\bibfnamefont {N.}~\bibnamefont
  {Moll}}, \bibinfo {author} {\bibfnamefont {Y.}~\bibnamefont {Xu}}, \bibinfo
  {author} {\bibfnamefont {O.~T.}\ \bibnamefont {Hofmann}}, \ and\ \bibinfo
  {author} {\bibfnamefont {P.}~\bibnamefont {Rinke}},\ }\bibfield  {title}
  {\enquote {\bibinfo {title} {Stabilization of semiconductor surfaces through
  bulk dopants},}\ }\href@noop {} {\bibfield  {journal} {\bibinfo  {journal}
  {New Journal of Physics}\ }\textbf {\bibinfo {volume} {15}},\ \bibinfo
  {pages} {083009} (\bibinfo {year} {2013})}\BibitemShut {NoStop}%
\bibitem [{\citenamefont {Sinai}\ \emph {et~al.}(2015)\citenamefont {Sinai},
  \citenamefont {Hofmann}, \citenamefont {Rinke}, \citenamefont {Scheffler},
  \citenamefont {Heimel},\ and\ \citenamefont {Kronik}}]{sinai2015multiscale}%
  \BibitemOpen
  \bibfield  {author} {\bibinfo {author} {\bibfnamefont {O.}~\bibnamefont
  {Sinai}}, \bibinfo {author} {\bibfnamefont {O.~T.}\ \bibnamefont {Hofmann}},
  \bibinfo {author} {\bibfnamefont {P.}~\bibnamefont {Rinke}}, \bibinfo
  {author} {\bibfnamefont {M.}~\bibnamefont {Scheffler}}, \bibinfo {author}
  {\bibfnamefont {G.}~\bibnamefont {Heimel}}, \ and\ \bibinfo {author}
  {\bibfnamefont {L.}~\bibnamefont {Kronik}},\ }\bibfield  {title} {\enquote
  {\bibinfo {title} {Multiscale approach to the electronic structure of doped
  semiconductor surfaces},}\ }\href@noop {} {\bibfield  {journal} {\bibinfo
  {journal} {Physical Review B}\ }\textbf {\bibinfo {volume} {91}},\ \bibinfo
  {pages} {075311} (\bibinfo {year} {2015})}\BibitemShut {NoStop}%
\bibitem [{\citenamefont {Fabiano}\ \emph {et~al.}(2009)\citenamefont
  {Fabiano}, \citenamefont {Piacenza}, \citenamefont {D’Agostino},\ and\
  \citenamefont {Della~Sala}}]{fabiano2009towards}%
  \BibitemOpen
  \bibfield  {author} {\bibinfo {author} {\bibfnamefont {E.}~\bibnamefont
  {Fabiano}}, \bibinfo {author} {\bibfnamefont {M.}~\bibnamefont {Piacenza}},
  \bibinfo {author} {\bibfnamefont {S.}~\bibnamefont {D’Agostino}}, \ and\
  \bibinfo {author} {\bibfnamefont {F.}~\bibnamefont {Della~Sala}},\ }\bibfield
   {title} {\enquote {\bibinfo {title} {Towards an accurate description of the
  electronic properties of the biphenylthiol/gold interface: The role of exact
  exchange},}\ }\href@noop {} {\bibfield  {journal} {\bibinfo  {journal} {The
  Journal of chemical physics}\ }\textbf {\bibinfo {volume} {131}},\ \bibinfo
  {pages} {234101} (\bibinfo {year} {2009})}\BibitemShut {NoStop}%
\bibitem [{\citenamefont {Biller}\ \emph {et~al.}(2011)\citenamefont {Biller},
  \citenamefont {Tamblyn}, \citenamefont {Neaton},\ and\ \citenamefont
  {Kronik}}]{biller2011electronic}%
  \BibitemOpen
  \bibfield  {author} {\bibinfo {author} {\bibfnamefont {A.}~\bibnamefont
  {Biller}}, \bibinfo {author} {\bibfnamefont {I.}~\bibnamefont {Tamblyn}},
  \bibinfo {author} {\bibfnamefont {J.~B.}\ \bibnamefont {Neaton}}, \ and\
  \bibinfo {author} {\bibfnamefont {L.}~\bibnamefont {Kronik}},\ }\bibfield
  {title} {\enquote {\bibinfo {title} {Electronic level alignment at a
  metal-molecule interface from a short-range hybrid functional},}\ }\href@noop
  {} {\bibfield  {journal} {\bibinfo  {journal} {The Journal of chemical
  physics}\ }\textbf {\bibinfo {volume} {135}},\ \bibinfo {pages} {164706}
  (\bibinfo {year} {2011})}\BibitemShut {NoStop}%
\bibitem [{\citenamefont {Hofmann}\ \emph {et~al.}(2013)\citenamefont
  {Hofmann}, \citenamefont {Atalla}, \citenamefont {Moll}, \citenamefont
  {Rinke},\ and\ \citenamefont {Scheffler}}]{hofmann2013interface}%
  \BibitemOpen
  \bibfield  {author} {\bibinfo {author} {\bibfnamefont {O.~T.}\ \bibnamefont
  {Hofmann}}, \bibinfo {author} {\bibfnamefont {V.}~\bibnamefont {Atalla}},
  \bibinfo {author} {\bibfnamefont {N.}~\bibnamefont {Moll}}, \bibinfo {author}
  {\bibfnamefont {P.}~\bibnamefont {Rinke}}, \ and\ \bibinfo {author}
  {\bibfnamefont {M.}~\bibnamefont {Scheffler}},\ }\bibfield  {title} {\enquote
  {\bibinfo {title} {Interface dipoles of organic molecules on ag (111) in
  hybrid density-functional theory},}\ }\href@noop {} {\bibfield  {journal}
  {\bibinfo  {journal} {New Journal of Physics}\ }\textbf {\bibinfo {volume}
  {15}},\ \bibinfo {pages} {123028} (\bibinfo {year} {2013})}\BibitemShut
  {NoStop}%
\bibitem [{\citenamefont {Wruss}\ \emph {et~al.}(2019)\citenamefont {Wruss},
  \citenamefont {Prokopiou}, \citenamefont {Kronik}, \citenamefont {Zojer},
  \citenamefont {Hofmann},\ and\ \citenamefont {Egger}}]{wruss2019magnetic}%
  \BibitemOpen
  \bibfield  {author} {\bibinfo {author} {\bibfnamefont {E.}~\bibnamefont
  {Wruss}}, \bibinfo {author} {\bibfnamefont {G.}~\bibnamefont {Prokopiou}},
  \bibinfo {author} {\bibfnamefont {L.}~\bibnamefont {Kronik}}, \bibinfo
  {author} {\bibfnamefont {E.}~\bibnamefont {Zojer}}, \bibinfo {author}
  {\bibfnamefont {O.~T.}\ \bibnamefont {Hofmann}}, \ and\ \bibinfo {author}
  {\bibfnamefont {D.~A.}\ \bibnamefont {Egger}},\ }\bibfield  {title} {\enquote
  {\bibinfo {title} {Magnetic configurations of open-shell molecules on metals:
  The case of cupc and copc on silver},}\ }\href@noop {} {\bibfield  {journal}
  {\bibinfo  {journal} {Physical Review Materials}\ }\textbf {\bibinfo {volume}
  {3}},\ \bibinfo {pages} {086002} (\bibinfo {year} {2019})}\BibitemShut
  {NoStop}%
\bibitem [{\citenamefont {Wruss}, \citenamefont {Zojer},\ and\ \citenamefont
  {Hofmann}(2018)}]{wruss2018distinguishing}%
  \BibitemOpen
  \bibfield  {author} {\bibinfo {author} {\bibfnamefont {E.}~\bibnamefont
  {Wruss}}, \bibinfo {author} {\bibfnamefont {E.}~\bibnamefont {Zojer}}, \ and\
  \bibinfo {author} {\bibfnamefont {O.~T.}\ \bibnamefont {Hofmann}},\
  }\bibfield  {title} {\enquote {\bibinfo {title} {Distinguishing between
  charge-transfer mechanisms at organic/inorganic interfaces employing hybrid
  functionals},}\ }\href@noop {} {\bibfield  {journal} {\bibinfo  {journal}
  {The Journal of Physical Chemistry C}\ }\textbf {\bibinfo {volume} {122}},\
  \bibinfo {pages} {14640--14653} (\bibinfo {year} {2018})}\BibitemShut
  {NoStop}%
\bibitem [{\citenamefont {Atalla}\ \emph {et~al.}(2013)\citenamefont {Atalla},
  \citenamefont {Yoon}, \citenamefont {Caruso}, \citenamefont {Rinke},\ and\
  \citenamefont {Scheffler}}]{atalla2013hybrid}%
  \BibitemOpen
  \bibfield  {author} {\bibinfo {author} {\bibfnamefont {V.}~\bibnamefont
  {Atalla}}, \bibinfo {author} {\bibfnamefont {M.}~\bibnamefont {Yoon}},
  \bibinfo {author} {\bibfnamefont {F.}~\bibnamefont {Caruso}}, \bibinfo
  {author} {\bibfnamefont {P.}~\bibnamefont {Rinke}}, \ and\ \bibinfo {author}
  {\bibfnamefont {M.}~\bibnamefont {Scheffler}},\ }\bibfield  {title} {\enquote
  {\bibinfo {title} {Hybrid density functional theory meets quasiparticle
  calculations: a consistent electronic structure approach},}\ }\href@noop {}
  {\bibfield  {journal} {\bibinfo  {journal} {Physical Review B}\ }\textbf
  {\bibinfo {volume} {88}},\ \bibinfo {pages} {165122} (\bibinfo {year}
  {2013})}\BibitemShut {NoStop}%
\bibitem [{\citenamefont {Paier}, \citenamefont {Marsman},\ and\ \citenamefont
  {Kresse}(2007)}]{paier2007does}%
  \BibitemOpen
  \bibfield  {author} {\bibinfo {author} {\bibfnamefont {J.}~\bibnamefont
  {Paier}}, \bibinfo {author} {\bibfnamefont {M.}~\bibnamefont {Marsman}}, \
  and\ \bibinfo {author} {\bibfnamefont {G.}~\bibnamefont {Kresse}},\
  }\bibfield  {title} {\enquote {\bibinfo {title} {Why does the b3lyp hybrid
  functional fail for metals?}}\ }\href@noop {} {\bibfield  {journal} {\bibinfo
   {journal} {The Journal of chemical physics}\ }\textbf {\bibinfo {volume}
  {127}},\ \bibinfo {pages} {024103} (\bibinfo {year} {2007})}\BibitemShut
  {NoStop}%
\bibitem [{\citenamefont {Wruss}, \citenamefont {H{\"o}rmann},\ and\
  \citenamefont {Hofmann}(2019)}]{wruss2019impact}%
  \BibitemOpen
  \bibfield  {author} {\bibinfo {author} {\bibfnamefont {E.}~\bibnamefont
  {Wruss}}, \bibinfo {author} {\bibfnamefont {L.}~\bibnamefont {H{\"o}rmann}},
  \ and\ \bibinfo {author} {\bibfnamefont {O.~T.}\ \bibnamefont {Hofmann}},\
  }\bibfield  {title} {\enquote {\bibinfo {title} {Impact of surface defects on
  the charge transfer at inorganic/organic interfaces},}\ }\href@noop {}
  {\bibfield  {journal} {\bibinfo  {journal} {The Journal of Physical Chemistry
  C}\ }\textbf {\bibinfo {volume} {123}},\ \bibinfo {pages} {7118--7124}
  (\bibinfo {year} {2019})}\BibitemShut {NoStop}%
\bibitem [{\citenamefont {Kresse}\ and\ \citenamefont
  {Hafner}(1993)}]{kresse1993ab}%
  \BibitemOpen
  \bibfield  {author} {\bibinfo {author} {\bibfnamefont {G.}~\bibnamefont
  {Kresse}}\ and\ \bibinfo {author} {\bibfnamefont {J.}~\bibnamefont
  {Hafner}},\ }\bibfield  {title} {\enquote {\bibinfo {title} {Ab initio
  molecular dynamics for liquid metals},}\ }\href@noop {} {\bibfield  {journal}
  {\bibinfo  {journal} {Physical Review B}\ }\textbf {\bibinfo {volume} {47}},\
  \bibinfo {pages} {558} (\bibinfo {year} {1993})}\BibitemShut {NoStop}%
\bibitem [{\citenamefont {Kresse}\ and\ \citenamefont
  {Hafner}(1994)}]{kresse1994ab}%
  \BibitemOpen
  \bibfield  {author} {\bibinfo {author} {\bibfnamefont {G.}~\bibnamefont
  {Kresse}}\ and\ \bibinfo {author} {\bibfnamefont {J.}~\bibnamefont
  {Hafner}},\ }\bibfield  {title} {\enquote {\bibinfo {title} {Ab initio
  molecular-dynamics simulation of the liquid-metal--amorphous-semiconductor
  transition in germanium},}\ }\href@noop {} {\bibfield  {journal} {\bibinfo
  {journal} {Physical Review B}\ }\textbf {\bibinfo {volume} {49}},\ \bibinfo
  {pages} {14251} (\bibinfo {year} {1994})}\BibitemShut {NoStop}%
\bibitem [{\citenamefont {Kresse}\ and\ \citenamefont
  {Furthm{\"u}ller}(1996{\natexlab{a}})}]{kresse1996efficiency}%
  \BibitemOpen
  \bibfield  {author} {\bibinfo {author} {\bibfnamefont {G.}~\bibnamefont
  {Kresse}}\ and\ \bibinfo {author} {\bibfnamefont {J.}~\bibnamefont
  {Furthm{\"u}ller}},\ }\bibfield  {title} {\enquote {\bibinfo {title}
  {Efficiency of ab-initio total energy calculations for metals and
  semiconductors using a plane-wave basis set},}\ }\href@noop {} {\bibfield
  {journal} {\bibinfo  {journal} {Computational materials science}\ }\textbf
  {\bibinfo {volume} {6}},\ \bibinfo {pages} {15--50} (\bibinfo {year}
  {1996}{\natexlab{a}})}\BibitemShut {NoStop}%
\bibitem [{\citenamefont {Kresse}\ and\ \citenamefont
  {Furthm{\"u}ller}(1996{\natexlab{b}})}]{kresse1996efficient}%
  \BibitemOpen
  \bibfield  {author} {\bibinfo {author} {\bibfnamefont {G.}~\bibnamefont
  {Kresse}}\ and\ \bibinfo {author} {\bibfnamefont {J.}~\bibnamefont
  {Furthm{\"u}ller}},\ }\bibfield  {title} {\enquote {\bibinfo {title}
  {Efficient iterative schemes for ab initio total-energy calculations using a
  plane-wave basis set},}\ }\href@noop {} {\bibfield  {journal} {\bibinfo
  {journal} {Physical review B}\ }\textbf {\bibinfo {volume} {54}},\ \bibinfo
  {pages} {11169} (\bibinfo {year} {1996}{\natexlab{b}})}\BibitemShut {NoStop}%
\bibitem [{\citenamefont {Bl{\"o}chl}(1994)}]{blochl1994projector}%
  \BibitemOpen
  \bibfield  {author} {\bibinfo {author} {\bibfnamefont {P.~E.}\ \bibnamefont
  {Bl{\"o}chl}},\ }\bibfield  {title} {\enquote {\bibinfo {title} {Projector
  augmented-wave method},}\ }\href@noop {} {\bibfield  {journal} {\bibinfo
  {journal} {Physical review B}\ }\textbf {\bibinfo {volume} {50}},\ \bibinfo
  {pages} {17953} (\bibinfo {year} {1994})}\BibitemShut {NoStop}%
\bibitem [{\citenamefont {Kresse}\ and\ \citenamefont
  {Joubert}(1999)}]{kresse1999ultrasoft}%
  \BibitemOpen
  \bibfield  {author} {\bibinfo {author} {\bibfnamefont {G.}~\bibnamefont
  {Kresse}}\ and\ \bibinfo {author} {\bibfnamefont {D.}~\bibnamefont
  {Joubert}},\ }\bibfield  {title} {\enquote {\bibinfo {title} {From ultrasoft
  pseudopotentials to the projector augmented-wave method},}\ }\href@noop {}
  {\bibfield  {journal} {\bibinfo  {journal} {Physical review b}\ }\textbf
  {\bibinfo {volume} {59}},\ \bibinfo {pages} {1758} (\bibinfo {year}
  {1999})}\BibitemShut {NoStop}%
\bibitem [{\citenamefont {Klime{\v{s}}}, \citenamefont {Bowler},\ and\
  \citenamefont {Michaelides}(2009)}]{klimevs2009chemical}%
  \BibitemOpen
  \bibfield  {author} {\bibinfo {author} {\bibfnamefont {J.}~\bibnamefont
  {Klime{\v{s}}}}, \bibinfo {author} {\bibfnamefont {D.~R.}\ \bibnamefont
  {Bowler}}, \ and\ \bibinfo {author} {\bibfnamefont {A.}~\bibnamefont
  {Michaelides}},\ }\bibfield  {title} {\enquote {\bibinfo {title} {Chemical
  accuracy for the van der waals density functional},}\ }\href@noop {}
  {\bibfield  {journal} {\bibinfo  {journal} {Journal of Physics: Condensed
  Matter}\ }\textbf {\bibinfo {volume} {22}},\ \bibinfo {pages} {022201}
  (\bibinfo {year} {2009})}\BibitemShut {NoStop}%
\bibitem [{\citenamefont {Klime{\v{s}}}, \citenamefont {Bowler},\ and\
  \citenamefont {Michaelides}(2011)}]{klimevs2011van}%
  \BibitemOpen
  \bibfield  {author} {\bibinfo {author} {\bibfnamefont {J.}~\bibnamefont
  {Klime{\v{s}}}}, \bibinfo {author} {\bibfnamefont {D.~R.}\ \bibnamefont
  {Bowler}}, \ and\ \bibinfo {author} {\bibfnamefont {A.}~\bibnamefont
  {Michaelides}},\ }\bibfield  {title} {\enquote {\bibinfo {title} {Van der
  waals density functionals applied to solids},}\ }\href@noop {} {\bibfield
  {journal} {\bibinfo  {journal} {Physical Review B}\ }\textbf {\bibinfo
  {volume} {83}},\ \bibinfo {pages} {195131} (\bibinfo {year}
  {2011})}\BibitemShut {NoStop}%
\bibitem [{\citenamefont {Peng}\ \emph {et~al.}(2016)\citenamefont {Peng},
  \citenamefont {Yang}, \citenamefont {Perdew},\ and\ \citenamefont
  {Sun}}]{peng2016versatile}%
  \BibitemOpen
  \bibfield  {author} {\bibinfo {author} {\bibfnamefont {H.}~\bibnamefont
  {Peng}}, \bibinfo {author} {\bibfnamefont {Z.-H.}\ \bibnamefont {Yang}},
  \bibinfo {author} {\bibfnamefont {J.~P.}\ \bibnamefont {Perdew}}, \ and\
  \bibinfo {author} {\bibfnamefont {J.}~\bibnamefont {Sun}},\ }\bibfield
  {title} {\enquote {\bibinfo {title} {Versatile van der waals density
  functional based on a meta-generalized gradient approximation},}\ }\href@noop
  {} {\bibfield  {journal} {\bibinfo  {journal} {Physical Review X}\ }\textbf
  {\bibinfo {volume} {6}},\ \bibinfo {pages} {041005} (\bibinfo {year}
  {2016})}\BibitemShut {NoStop}%
\bibitem [{\citenamefont {Henze}\ \emph {et~al.}(2007)\citenamefont {Henze},
  \citenamefont {Bauer}, \citenamefont {Lee}, \citenamefont {Sokolowski},\ and\
  \citenamefont {Tautz}}]{henze2007vertical}%
  \BibitemOpen
  \bibfield  {author} {\bibinfo {author} {\bibfnamefont {S.}~\bibnamefont
  {Henze}}, \bibinfo {author} {\bibfnamefont {O.}~\bibnamefont {Bauer}},
  \bibinfo {author} {\bibfnamefont {T.-L.}\ \bibnamefont {Lee}}, \bibinfo
  {author} {\bibfnamefont {M.}~\bibnamefont {Sokolowski}}, \ and\ \bibinfo
  {author} {\bibfnamefont {F.}~\bibnamefont {Tautz}},\ }\bibfield  {title}
  {\enquote {\bibinfo {title} {Vertical bonding distances of ptcda on au (1 1
  1) and ag (1 1 1): Relation to the bonding type},}\ }\href@noop {} {\bibfield
   {journal} {\bibinfo  {journal} {Surface Science}\ }\textbf {\bibinfo
  {volume} {601}},\ \bibinfo {pages} {1566--1573} (\bibinfo {year}
  {2007})}\BibitemShut {NoStop}%
\bibitem [{\citenamefont {Todorovi{\'c}}\ \emph {et~al.}(2019)\citenamefont
  {Todorovi{\'c}}, \citenamefont {Gutmann}, \citenamefont {Corander},\ and\
  \citenamefont {Rinke}}]{todorovic2019bayesian}%
  \BibitemOpen
  \bibfield  {author} {\bibinfo {author} {\bibfnamefont {M.}~\bibnamefont
  {Todorovi{\'c}}}, \bibinfo {author} {\bibfnamefont {M.~U.}\ \bibnamefont
  {Gutmann}}, \bibinfo {author} {\bibfnamefont {J.}~\bibnamefont {Corander}}, \
  and\ \bibinfo {author} {\bibfnamefont {P.}~\bibnamefont {Rinke}},\ }\bibfield
   {title} {\enquote {\bibinfo {title} {Bayesian inference of atomistic
  structure in functional materials},}\ }\href@noop {} {\bibfield  {journal}
  {\bibinfo  {journal} {npj Computational Materials}\ }\textbf {\bibinfo
  {volume} {5}},\ \bibinfo {pages} {1--7} (\bibinfo {year} {2019})}\BibitemShut
  {NoStop}%
\bibitem [{\citenamefont {Egger}\ \emph {et~al.}()\citenamefont {Egger},
  \citenamefont {Hörmann}, \citenamefont {Jeindl}, \citenamefont {Scherbela},
  \citenamefont {Obersteiner}, \citenamefont {Todorović}, \citenamefont
  {Rinke},\ and\ \citenamefont {Hofmann}}]{doi:10.1002/advs.202000992}%
  \BibitemOpen
  \bibfield  {author} {\bibinfo {author} {\bibfnamefont {A.~T.}\ \bibnamefont
  {Egger}}, \bibinfo {author} {\bibfnamefont {L.}~\bibnamefont {Hörmann}},
  \bibinfo {author} {\bibfnamefont {A.}~\bibnamefont {Jeindl}}, \bibinfo
  {author} {\bibfnamefont {M.}~\bibnamefont {Scherbela}}, \bibinfo {author}
  {\bibfnamefont {V.}~\bibnamefont {Obersteiner}}, \bibinfo {author}
  {\bibfnamefont {M.}~\bibnamefont {Todorović}}, \bibinfo {author}
  {\bibfnamefont {P.}~\bibnamefont {Rinke}}, \ and\ \bibinfo {author}
  {\bibfnamefont {O.~T.}\ \bibnamefont {Hofmann}},\ }\bibfield  {title}
  {\enquote {\bibinfo {title} {Charge transfer into organic thin films: A
  deeper insight through machine-learning-assisted structure search},}\ }\href
  {\doibase 10.1002/advs.202000992} {\bibfield  {journal} {\bibinfo  {journal}
  {Advanced Science}\ }\textbf {\bibinfo {volume} {n/a}},\ \bibinfo {pages}
  {2000992}},\ \Eprint
  {http://arxiv.org/abs/https://onlinelibrary.wiley.com/doi/pdf/10.1002/advs.202000992}
  {https://onlinelibrary.wiley.com/doi/pdf/10.1002/advs.202000992} \BibitemShut
  {NoStop}%
\bibitem [{\citenamefont {Jure{\v{c}}ka}\ \emph {et~al.}(2006)\citenamefont
  {Jure{\v{c}}ka}, \citenamefont {{\v{S}}poner}, \citenamefont
  {{\v{C}}ern{\`y}},\ and\ \citenamefont {Hobza}}]{jurevcka2006benchmark}%
  \BibitemOpen
  \bibfield  {author} {\bibinfo {author} {\bibfnamefont {P.}~\bibnamefont
  {Jure{\v{c}}ka}}, \bibinfo {author} {\bibfnamefont {J.}~\bibnamefont
  {{\v{S}}poner}}, \bibinfo {author} {\bibfnamefont {J.}~\bibnamefont
  {{\v{C}}ern{\`y}}}, \ and\ \bibinfo {author} {\bibfnamefont {P.}~\bibnamefont
  {Hobza}},\ }\bibfield  {title} {\enquote {\bibinfo {title} {Benchmark
  database of accurate (mp2 and ccsd (t) complete basis set limit) interaction
  energies of small model complexes, dna base pairs, and amino acid pairs},}\
  }\href@noop {} {\bibfield  {journal} {\bibinfo  {journal} {Physical Chemistry
  Chemical Physics}\ }\textbf {\bibinfo {volume} {8}},\ \bibinfo {pages}
  {1985--1993} (\bibinfo {year} {2006})}\BibitemShut {NoStop}%
\bibitem [{\citenamefont {Puzder}, \citenamefont {Dion},\ and\ \citenamefont
  {Langreth}(2006)}]{puzder2006binding}%
  \BibitemOpen
  \bibfield  {author} {\bibinfo {author} {\bibfnamefont {A.}~\bibnamefont
  {Puzder}}, \bibinfo {author} {\bibfnamefont {M.}~\bibnamefont {Dion}}, \ and\
  \bibinfo {author} {\bibfnamefont {D.~C.}\ \bibnamefont {Langreth}},\
  }\bibfield  {title} {\enquote {\bibinfo {title} {Binding energies in benzene
  dimers: Nonlocal density functional calculations},}\ }\href@noop {}
  {\bibfield  {journal} {\bibinfo  {journal} {The Journal of chemical physics}\
  }\textbf {\bibinfo {volume} {124}},\ \bibinfo {pages} {164105} (\bibinfo
  {year} {2006})}\BibitemShut {NoStop}%
\bibitem [{\citenamefont {Toyoda}\ \emph {et~al.}(2009)\citenamefont {Toyoda},
  \citenamefont {Nakano}, \citenamefont {Hamada}, \citenamefont {Lee},
  \citenamefont {Yanagisawa},\ and\ \citenamefont
  {Morikawa}}]{toyoda2009first}%
  \BibitemOpen
  \bibfield  {author} {\bibinfo {author} {\bibfnamefont {K.}~\bibnamefont
  {Toyoda}}, \bibinfo {author} {\bibfnamefont {Y.}~\bibnamefont {Nakano}},
  \bibinfo {author} {\bibfnamefont {I.}~\bibnamefont {Hamada}}, \bibinfo
  {author} {\bibfnamefont {K.}~\bibnamefont {Lee}}, \bibinfo {author}
  {\bibfnamefont {S.}~\bibnamefont {Yanagisawa}}, \ and\ \bibinfo {author}
  {\bibfnamefont {Y.}~\bibnamefont {Morikawa}},\ }\bibfield  {title} {\enquote
  {\bibinfo {title} {First-principles study of benzene on noble metal surfaces:
  Adsorption states and vacuum level shifts},}\ }\href@noop {} {\bibfield
  {journal} {\bibinfo  {journal} {Surface science}\ }\textbf {\bibinfo {volume}
  {603}},\ \bibinfo {pages} {2912--2922} (\bibinfo {year} {2009})}\BibitemShut
  {NoStop}%
\bibitem [{\citenamefont {Lovinger}\ \emph {et~al.}(1984)\citenamefont
  {Lovinger}, \citenamefont {Forrest}, \citenamefont {Kaplan}, \citenamefont
  {Schmidt},\ and\ \citenamefont {Venkatesan}}]{lovinger1984structural}%
  \BibitemOpen
  \bibfield  {author} {\bibinfo {author} {\bibfnamefont {A.}~\bibnamefont
  {Lovinger}}, \bibinfo {author} {\bibfnamefont {S.}~\bibnamefont {Forrest}},
  \bibinfo {author} {\bibfnamefont {M.}~\bibnamefont {Kaplan}}, \bibinfo
  {author} {\bibfnamefont {P.}~\bibnamefont {Schmidt}}, \ and\ \bibinfo
  {author} {\bibfnamefont {T.}~\bibnamefont {Venkatesan}},\ }\bibfield  {title}
  {\enquote {\bibinfo {title} {Structural and morphological investigation of
  the development of electrical conductivity in ion-irradiated thin films of an
  organic material},}\ }\href@noop {} {\bibfield  {journal} {\bibinfo
  {journal} {Journal of applied physics}\ }\textbf {\bibinfo {volume} {55}},\
  \bibinfo {pages} {476--482} (\bibinfo {year} {1984})}\BibitemShut {NoStop}%
\bibitem [{\citenamefont {M{\"o}bus}, \citenamefont {Karl},\ and\ \citenamefont
  {Kobayashi}(1992)}]{mobus1992structure}%
  \BibitemOpen
  \bibfield  {author} {\bibinfo {author} {\bibfnamefont {M.}~\bibnamefont
  {M{\"o}bus}}, \bibinfo {author} {\bibfnamefont {N.}~\bibnamefont {Karl}}, \
  and\ \bibinfo {author} {\bibfnamefont {T.}~\bibnamefont {Kobayashi}},\
  }\bibfield  {title} {\enquote {\bibinfo {title} {Structure of
  perylene-tetracarboxylic-dianhydride thin films on alkali halide crystal
  substrates},}\ }\href@noop {} {\bibfield  {journal} {\bibinfo  {journal}
  {Journal of crystal growth}\ }\textbf {\bibinfo {volume} {116}},\ \bibinfo
  {pages} {495--504} (\bibinfo {year} {1992})}\BibitemShut {NoStop}%
\bibitem [{\citenamefont {Ogawa}\ \emph {et~al.}(1999)\citenamefont {Ogawa},
  \citenamefont {Kuwamoto}, \citenamefont {Isoda}, \citenamefont {Kobayashi},\
  and\ \citenamefont {Karl}}]{ogawa19993}%
  \BibitemOpen
  \bibfield  {author} {\bibinfo {author} {\bibfnamefont {T.}~\bibnamefont
  {Ogawa}}, \bibinfo {author} {\bibfnamefont {K.}~\bibnamefont {Kuwamoto}},
  \bibinfo {author} {\bibfnamefont {S.}~\bibnamefont {Isoda}}, \bibinfo
  {author} {\bibfnamefont {T.}~\bibnamefont {Kobayashi}}, \ and\ \bibinfo
  {author} {\bibfnamefont {N.}~\bibnamefont {Karl}},\ }\bibfield  {title}
  {\enquote {\bibinfo {title} {3, 4: 9, 10-perylenetetracarboxylic dianhydride
  (ptcda) by electron crystallography},}\ }\href@noop {} {\bibfield  {journal}
  {\bibinfo  {journal} {Acta Crystallographica Section B: Structural Science}\
  }\textbf {\bibinfo {volume} {55}},\ \bibinfo {pages} {123--130} (\bibinfo
  {year} {1999})}\BibitemShut {NoStop}%
\bibitem [{\citenamefont {Mannsfeld}\ and\ \citenamefont
  {Fritz}(2004)}]{mannsfeld2004analysis}%
  \BibitemOpen
  \bibfield  {author} {\bibinfo {author} {\bibfnamefont {S.~B.}\ \bibnamefont
  {Mannsfeld}}\ and\ \bibinfo {author} {\bibfnamefont {T.}~\bibnamefont
  {Fritz}},\ }\bibfield  {title} {\enquote {\bibinfo {title} {Analysis of the
  substrate influence on the ordering of epitaxial molecular layers: The
  special case of point-on-line coincidence},}\ }\href@noop {} {\bibfield
  {journal} {\bibinfo  {journal} {Physical Review B}\ }\textbf {\bibinfo
  {volume} {69}},\ \bibinfo {pages} {075416} (\bibinfo {year}
  {2004})}\BibitemShut {NoStop}%
\bibitem [{\citenamefont {Mannsfeld}\ and\ \citenamefont
  {Fritz}(2006)}]{mannsfeld2006advanced}%
  \BibitemOpen
  \bibfield  {author} {\bibinfo {author} {\bibfnamefont {S.~C.}\ \bibnamefont
  {Mannsfeld}}\ and\ \bibinfo {author} {\bibfnamefont {T.}~\bibnamefont
  {Fritz}},\ }\bibfield  {title} {\enquote {\bibinfo {title} {Advanced
  modelling of epitaxial ordering of organic layers on crystalline surfaces},}\
  }\href@noop {} {\bibfield  {journal} {\bibinfo  {journal} {Modern Physics
  Letters B}\ }\textbf {\bibinfo {volume} {20}},\ \bibinfo {pages} {585--605}
  (\bibinfo {year} {2006})}\BibitemShut {NoStop}%
\bibitem [{\citenamefont {Hörmann}, \citenamefont {Jeindl},\ and\
  \citenamefont {Hofmann}(2020)}]{PTCDA_on_Ag111_data}%
  \BibitemOpen
  \bibfield  {author} {\bibinfo {author} {\bibfnamefont {L.}~\bibnamefont
  {Hörmann}}, \bibinfo {author} {\bibfnamefont {A.}~\bibnamefont {Jeindl}}, \
  and\ \bibinfo {author} {\bibfnamefont {O.~T.}\ \bibnamefont {Hofmann}},\
  }\href {\doibase 10.17172/NOMAD/2020.08.16-1} {\enquote {\bibinfo {title}
  {Data for ptcda on ag(111)},}\ }\bibinfo {howpublished}
  {\url{https://dx.doi.org/10.17172/NOMAD/2020.08.16-1}} (\bibinfo {year}
  {2020})\BibitemShut {NoStop}%
\end{thebibliography}%

\end{document}